\documentclass[pra,aps,reprint,a4paper,superscriptaddress,longbibliography,floatfix]{revtex4-1}

\usepackage{bm}
\usepackage{times}
\usepackage{epsfig}
\usepackage{amsfonts}
\usepackage{amsmath}
\usepackage{amssymb}
\usepackage{amsthm}
\usepackage{color}
\usepackage{multirow}
\usepackage[colorlinks=true,linkcolor=blue,citecolor=magenta,urlcolor=blue]{hyperref}
\usepackage{amsthm}
\usepackage{epigraph}
\usepackage{thmtools}
\usepackage{thm-restate}

\newcommand{\thb}[0]{\operatorname{TH}}

\newcommand{\qstab}[0]{\operatorname{QSTAB}}
\newcommand{\abl}[0]{\operatorname{abl}}
\newcommand{\id}[0]{\mathbb{I}}

\begin{document}

%%%%%%%%%%%%%%%%%%%%%%%%%%%%%%%%%%%%%%%%%%%%%%%%%%%%%%%%%%%%%%%%%%%

\title{Quantum correlations from simple assumptions}

%%%%%%%%%%%%%%%%%%%%%%%%%%%%%%%%%%%%%%%%%%%%%%%%%%%%%%%%%%%%%%%%%%%

\author{Ad\'an Cabello}
\email{adan@us.es}
\affiliation{Departamento de F\'{\i}sica Aplicada II, Universidad de
Sevilla, E-41012 Sevilla, Spain}
\affiliation{Instituto Carlos~I de F\'{\i}sica Te\'orica y Computacional, Universidad de
Sevilla, E-41012 Sevilla, Spain}

%%%%%%%%%%%%%%%%%%%%%%%%%%%%%%%%%%%%%%%%%%%%%%%%%%%%%%%%%%%%%%%%%%%

\date{\today}

%%%%%%%%%%%%%%%%%%%%%%%%%%%%%%%%%%%%%%%%%%%%%%%%%%%%%%%%%%%%%%%%%%%

\begin{abstract}
We address the problem of deriving the set of quantum correlations for every Bell and Kochen-Specker (KS) contextuality scenario from simple assumptions. We show that the correlations that are possible according to quantum theory are equal to those possible under the assumptions that there is a nonempty set of correlations for every KS scenario and a statistically independent realization of any two KS experiments. The proof uses tools of the graph-theoretic approach to correlations and deals with Bell nonlocality and KS contextuality in a unified way.
\end{abstract} 

%%%%%%%%%%%%%%%%%%%%%%%%%%%%%%%%%%%%%%%%%%%%%%%%%%%%%%%%%%%%%%%%%%%

\maketitle

%%%%%%%%%%%%%%%%%%%%%%%%%%%%%%%%%%%%%%%%%%%%%%%%%%%%%%%%%%%%%%%%%%%

\section{Introduction}

%%%%%%%%%%%%%%%%%%%%%%%%%%%%%%%%%%%%%%%%%%%%%%%%%%%%%%%%%%%%%%%%%%%

The problem of whether quantum theory (QT) allowed an explanation in terms of hidden variables arose at the precise instant that Born proposed the probabilistic interpretation of the wave function \cite{Born26}. The two most famous results on the problem of hidden variables in QT are two theorems. Bell's theorem \cite{Bell64,CHSH69} asserts that it is impossible to explain, using local hidden-variable theories, the quantum correlations arising in experiments in which two or more spatially separated parties perform measurements on different subsystems of a composite system.
The Kochen-Specker (KS) theorem \cite{Specker60,Bell66,KS67} states that it is impossible to explain the predictions of QT for ideal measurements assuming that they reveal preexisting results that are independent on which other compatible ideal measurements are carried out. The theories in which this is the case are called KS noncontextual hidden-variable theories.

The proof of Bell's theorem is based on identifying quantum correlations that violate inequalities that must be satisfied by local hidden-variables theories. These inequalities are called Bell's inequalities and their violation is called Bell nonlocality \cite{BCPSW14}. It took a surprising amount of time until someone asked a natural question: what is the maximum violation of Bell's inequalities that QT permits? \cite{Tsirelson80}. It took even more time until someone asked another natural question: what prevents larger violations? \cite{PR94}. It was surprising to discover that the no-signaling principle allows for larger-than-quantum violations of Bell's inequalities \cite{PR94}. The answer was sought examining the implications of Bell nonlocality for distributed computation \cite{vanDam99,LPSW07}. Then, two different principles, information causality \cite{PPKSWZ09} and macroscopic locality \cite{NW09}, were able to account for the maximum quantum violation of the simplest Bell inequality. However, it was soon found that the problem was more complex as there were correlations that do not violate Bell's inequalities beyond the maximum quantum bound but are still forbidden by QT, although they seem to satisfy all proposed principles \cite{NGHA15}.

In parallel, it was found that the KS theorem can be proven using Bell-like inequalities involving the correlations between the outcomes of sequential compatible ideal measurements \cite{KCBS08,Cabello08}. These inequalities are called KS noncontextuality (NC) inequalities and their violation is called KS contextuality. Bell inequalities with ideal measurements are KS NC inequalities. Also in the case of KS NC inequalities, it was found that larger than quantum violations were possible, raising the question of what prevents nonquantum correlations \cite{CSW10}. It was proven that the so-called exclusivity principle explains the maximum quantum violation of the simplest KS NC inequality violated by the simplest quantum systems \cite{Cabello13}. 

If we combine both perspectives, we can ask what is the principle that explains why, according to QT, some forms of Bell nonlocality and KS contextuality are possible while others are not.
The aim of this article is to address this problem and show that the perspective that unifies Bell nonlocality and KS contextuality is actually useful to solve it. The solution we propose benefits from this unifying perspective and uses tools of the so-called graph-theoretic approach to quantum correlations introduced in Refs.~\cite{CSW10,CSW14}. The article is, however, self-contained and does not require previous knowledge of the graph-theoretic approach. 

The structure of the article is the following. In Sec.~\ref{Sec2} we define Bell and KS contextuality scenarios, explain what is meant by ``correlations,'' and recall which are the correlations that are physically possible according to QT. In Sec.~\ref{Sec3} we present our result: a derivation of the quantum correlations for Bell and KS contextuality scenarios based on two assumptions. The proof is developed in Sec.~\ref{Sec4}. In Sec.~\ref{Sec5} we present our conclusions. In addition, we include in two appendixes the proofs of two lemmas used in Sec.~\ref{Sec4}. 

%%%%%%%%%%%%%%%%%%%%%%%%%%%%%%%%%%%%%%%%%%%%%%%%%%%%%%%%%%%%%%%%%%%

\section{Quantum correlations in Bell and KS scenarios}
\label{Sec2}

%%%%%%%%%%%%%%%%%%%%%%%%%%%%%%%%%%%%%%%%%%%%%%%%%%%%%%%%%%%%%%%%%%%

\subsection{Compatibility, nondisturbance, and ideal measurements}
\label{sus}

%%%%%%%%%%%%%%%%%%%%%%%%%%%%%%%%%%%%%%%%%%%%%%%%%%%%%%%%%%%%%%%%%%%

In this article we consider physical theories that assign probabilities to the outcomes of measurements. We will use $P(x=a | \psi)$ to denote the probability of outcome $a$ after measuring $x$ on state~$\psi$. By ``state'' we mean the object that encodes the expectations about the outcomes of future measurements. We do not assume any particular mathematical representation for the states, measurements, and outcomes.

{\em Definition 1.}
	A measurement $z$ with outcomes $c \in C$ is a coarse graining of a measurement $x$ with outcomes $a \in A$ if, for all $c \in C$, there is $A_c \subseteq A$ such that, for all states $\psi$, 
	\begin{equation}
	P(z=c | \psi) = \sum_{a \in A_c} P(x=a | \psi)
	\end{equation}
	and $A_c \cap A_{c'} = \emptyset$ if $c \neq c'$.

{\em Definition 2.}
	Two measurements are compatible if they are coarse grainings of the same measurement.

{\em Definition 3.}
	Two sets of measurements, $X=\{x_i\}$, with respective outcomes $a_i \in A_i$, and $Y=\{y_j\}$, with respective outcomes $b_j \in B_j$, such that every pair $(x_i,y_j)$ are compatible, are mutually nondisturbing if, for all $x_i \in X$, $a_i \in A_i$, and $y_j,y_k \in Y$,
	\begin{equation}
	\label{no-disturbance1}
	\sum_{b_j \in B_j} P(x_i=a_i,y_j=b_j|\psi)= \sum_{b_k \in B_k} P(x_i=a_i,y_k=b_k|\psi), 
	\end{equation}
	and, for all $y_i \in Y$, $b_i \in B_i$, and $x_j,x_k \in X$,
	\begin{equation}
	\label{no-disturbance2}
	\sum_{a_j \in A_j} P(x_j=a_j,y_i=b_i|\psi)= \sum_{a_k \in A_k} P(x_k=a_k,y_i=b_i|\psi).
	\end{equation} 
	Therefore, the marginal probabilities $P(x_i=a_i|\psi)$ are independent of the choice of measurement $y_j \in Y_j$ and the marginal probabilities $P(y_i=b_i|\psi)$ are independent of the choice of measurement $x_j \in X_j$.

{\em Definition 4.}
	A measurement is ideal (or sharp \cite{CY14}; see also \cite{Kleinmann14,CY16}) if (i) it gives the same outcome when performed consecutive times on the same physical system, (ii) it does not disturb compatible measurements, and (iii) all its coarse grainings satisfy (i) and (ii).

%%%%%%%%%%%%%%%%%%%%%%%%%%%%%%%%%%%%%%%%%%%%%%%%%%%%%%%%%%%%%%%%%%%

\subsection{Bell scenarios}

%%%%%%%%%%%%%%%%%%%%%%%%%%%%%%%%%%%%%%%%%%%%%%%%%%%%%%%%%%%%%%%%%%%

In a Bell nonlocality experiment \cite{Bell64,CHSH69} there are two (or more) spatially separated parties, typically called Alice and Bob. Each of them acts on a different subsystem of a composite system. In each round of the experiment, Alice performs a freely chosen measurement $x \in X$ and Bob performs a freely chosen measurement~$y \in Y$. The two measurements are spacelike separated in the sense that the region of space-time from one party's choice to the recording of its outcome is spacelike separated from the corresponding region for the other party. Therefore, assuming that faster-than-light communication is impossible, measurements performed by different parties are mutually nondisturbing. 

Each Bell scenario is characterized by the number of parties, the number of measurements each party can perform, the number of outcomes of each measurement, and the relations of compatibility between the measurements.
For example, the simplest Bell scenario is the Clauser-Horne-Shimony-Holt (CHSH) or $(2,2,2)$ Bell scenario \cite{Bell64,CHSH69} in which there are two parties, each of them can perform one out of two measurements, each of them has two outcomes, and every measurement of Alice is compatible with every measurement of Bob. 

%%%%%%%%%%%%%%%%%%%%%%%%%%%%%%%%%%%%%%%%%%%%%%%%%%%%%%%%%%%%%%%%%%%

\subsection{KS contextuality scenarios}
\label{ide}

%%%%%%%%%%%%%%%%%%%%%%%%%%%%%%%%%%%%%%%%%%%%%%%%%%%%%%%%%%%%%%%%%%%

KS contextuality scenarios (hereafter KS scenarios for brevity) extend Bell scenarios to cover situations in which compatible measurements are not necessarily spacelike separated. As in the case of Bell scenarios, a KS scenario is characterized by a number of measurements (each of them with a number of outcomes) and their relations of compatibility. However, in KS scenarios measurements are assumed to be ideal. The restriction to ideal measurements (restriction that does not exist in Bell scenarios) makes compatible measurements automatically mutually nondisturbing (as occurs in Bell scenarios). Any Bell scenario with ideal measurements is a KS scenario, but Bell scenarios with nonideal measurements are not KS scenarios. However, as we shall see in Sec.~\ref{qc}, all quantum correlations that can be attained in Bell scenarios with nonideal measurements can also be attained with ideal measurements.

In classical physics, ideal measurements reveal preexisting outcomes that are independent of which other compatible measurements are performed. Therefore, if by ``context'' we mean a set of compatible measurements, it is reasonable to make the assumption that the hypothetical hidden variable explanations of QT satisfy, for ideal measurements, that measurement outcomes are independent of the context~\cite{Specker60,Bell66,KS67,KCBS08,Cabello08}. However, this assumption is, in general, not justified if we remove the restriction to ideal measurements \cite{Spekkens14}.

The simplest KS scenario in which qutrits (which are the simplest quantum systems producing KS contextuality) produce KS contextuality is the Klyachko-Can-Binicio\u{g}lu-Shumovsky (KCBS) KS scenario \cite{KCBS08}, involving five measurements $x_i$, with $i=1,\ldots,5$ (with two possible outcomes), such that $x_i$ and $x_{i + 1}$ (with the sum taken modulo five) are compatible.

%%%%%%%%%%%%%%%%%%%%%%%%%%%%%%%%%%%%%%%%%%%%%%%%%%%%%%%%%%%%%%%%%%%

\subsection{Contexts and graphs of compatibility}

%%%%%%%%%%%%%%%%%%%%%%%%%%%%%%%%%%%%%%%%%%%%%%%%%%%%%%%%%%%%%%%%%%%

{\em Definition 5.}
	A context in a Bell or KS scenario is a subset of compatible (and mutually nondisturbing) measurements. 

Any subset of a context is also a context.

The relations of compatibility between the measurements in a Bell or KS scenario~$S$ can be represented by a graph (see, e.g., Refs.~\cite{KRK12,CDLP13}) in which vertices represent measurements and edges relations of compatibility. A graph with this interpretation is called the graph of compatibility (or compatibility graph) of~$S$. 

In a graph of compatibility, contexts are represented by cliques. A clique is a set of vertices every pair of which are adjacent. For example, the graph of compatibility of CHSH Bell scenario \cite{Bell64,CHSH69} is a square (and has four cliques of size two) and the graph of compatibility of the KCBS KS scenario \cite{KCBS08} is a pentagon (and has five cliques of size two).

%%%%%%%%%%%%%%%%%%%%%%%%%%%%%%%%%%%%%%%%%%%%%%%%%%%%%%%%%%%%%%%%%%%

\subsection{Mutually exclusive events and graphs of exclusivity}

%%%%%%%%%%%%%%%%%%%%%%%%%%%%%%%%%%%%%%%%%%%%%%%%%%%%%%%%%%%%%%%%%%%

The possible events of scenario $S$ correspond to all possible combinations of outcomes of the compatible measurements in $S$. For example, in the CHSH Bell scenario \cite{Bell64,CHSH69}, there are 16~events $(x=a,y=b|\psi)$ with $x,y,a,b \in \{0,1\}$. In some places, for brevity, we will use $(ab|xy)$ to represent $(x=a,y=b|\psi)$.

{\em Definition 6.}
Two events of $S$ are mutually exclusive if there is a measurement $M$ such that each event corresponds to a different outcome of $M$. 

Measurement $M$ must be constructed using the measurements, outcomes, and compatibility relations available in~$S$. For example, in the CHSH Bell scenario, $(ab|xy)$ and $(a'b'|x'y')$ are mutually exclusive in the following cases. 

(i) When $x=x'$, $a \neq a'$, $y = y'$, and $b \neq b'$. Then, $M$ is the four-outcome measurement in which Alice measures $x$ and Bob measures $y$. That is, the measurement that produces events $(00|xy)$, $(01|xy)$, $(10|xy)$, and $(11|xy)$.

(ii) When $x=x'$, $a \neq a'$, and $y \neq y'$. Then, $M$ is the measurement that produces events $(a0|xy)$, $(a1|xy)$, $(a'0|xy')$, and $(a'1|xy')$. This $M$ is defined by suitably selecting Alice's and Bob's events. By definition of Bell scenario there is no causal relation between one party's choice and the other party's outcome.

(iii) When $x \neq x'$, $y = y'$, and $b \neq b'$. Then, $M$ is the measurement that produces events $(0b|xy)$, $(1b|xy)$, $(0b'|x'y)$, and $(1b'|x'y)$. This $M$ is defined by suitably selecting Alice's and Bob's events. By definition of Bell scenario there is no causal relation between one party's choice and the other party's outcome.

The relations of mutual exclusivity between events can be represented by a graph \cite{CSW10,CSW14} in which vertices represent events and edges represent relations of mutual exclusivity. A graph with this interpretation is called a graph of exclusivity (or exclusivity graph).

%%%%%%%%%%%%%%%%%%%%%%%%%%%%%%%%%%%%%%%%%%%%%%%%%%%%%%%%%%%%%%%%%%%
% Fig. 1
%%%%%%%%%%%%%%%%%%%%%%%%%%%%%%%%%%%%%%%%%%%%%%%%%%%%%%%%%%%%%%%%%%%

\begin{figure}[t]
	\vspace{-9mm}
	\hspace{-16mm}
	\includegraphics[width=10.1cm]{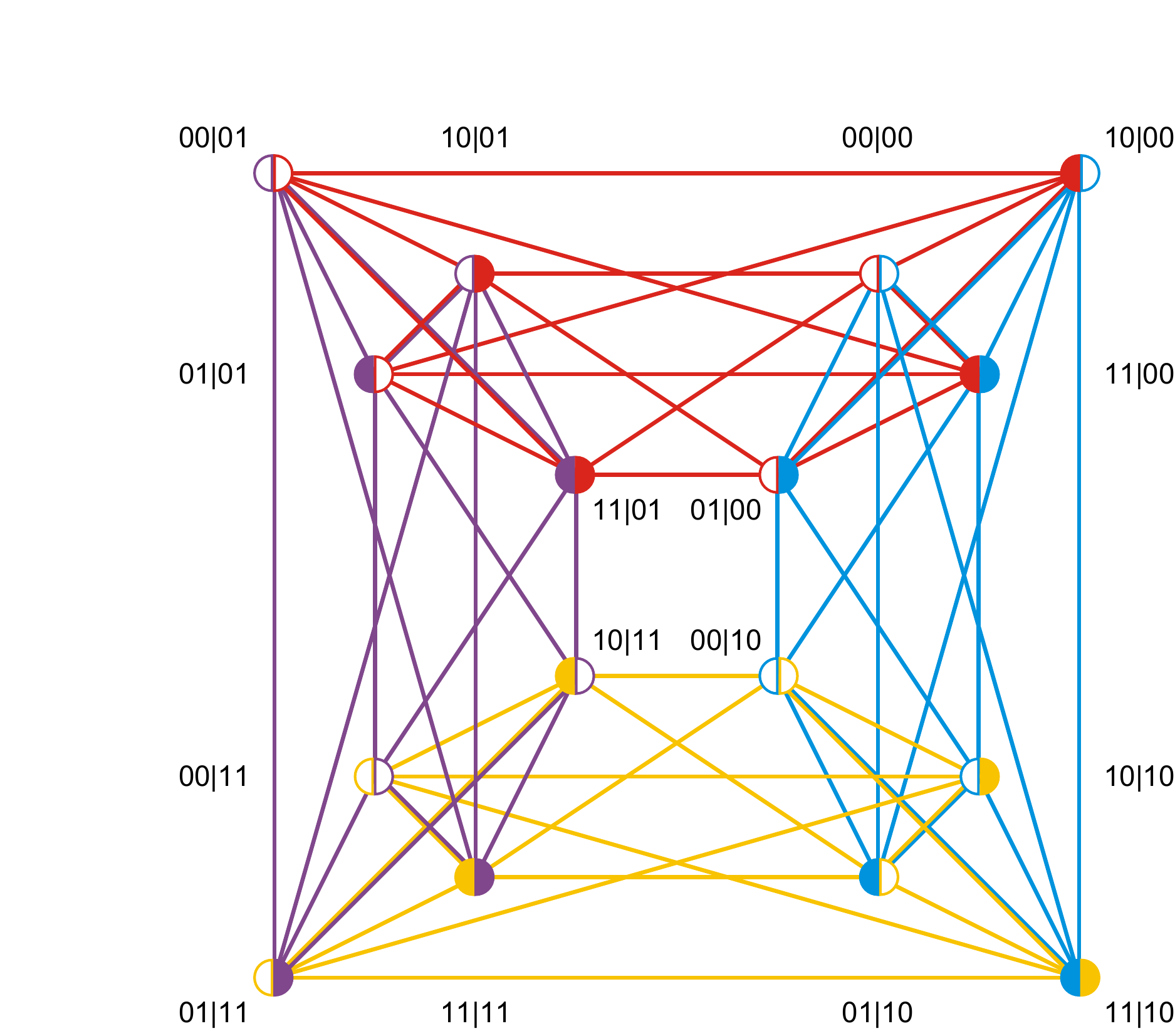}
	\caption{Graph of exclusivity of the $16$~elementary events of the CHSH Bell scenario. Each node represents an event $(x=a,y=b|\psi)$ denoted for brevity $ab|xy$. Each color represents one measurement: Red and yellow for Alice's measurements zero and $1$, respectively. Cyan and purple for Bob's measurements zero and $1$, respectively. Each event is defined by the outcomes of two compatible measurements, one of Alice and one of Bob. This is represented by dividing each node in two semicircles. An empty (full) semicircle indicates that the outcome of the corresponding measurement is zero (respectively, $1$).}
	\label{Fig1}
\end{figure}

%%%%%%%%%%%%%%%%%%%%%%%%%%%%%%%%%%%%%%%%%%%%%%%%%%%%%%%%%%%%%%%%%%%

Given a set of events of a Bell or KS scenario~$S$, the measurements and outcomes that define each event and the measurements that made some of the pairs of events mutually exclusive can be represented within the graph of exclusivity by representing each measurement in~$S$ by a different color and each outcome by a different symbol. For the $16$~elementary events of the CHSH Bell scenario this graph of exclusivity is shown in Fig.~\ref{Fig1}.

%%%%%%%%%%%%%%%%%%%%%%%%%%%%%%%%%%%%%%%%%%%%%%%%%%%%%%%%%%%%%%%%%%%

\subsection{Correlations}
\label{corre}

%%%%%%%%%%%%%%%%%%%%%%%%%%%%%%%%%%%%%%%%%%%%%%%%%%%%%%%%%%%%%%%%%%%

What we informally refer to as ``correlations'' for a particular Bell or KS scenario~$S$ is a set $\mathbf{p}(S) \in \cal {P}(S)$ of probability distributions, one for each context. Following the terminology introduced in Ref.~\cite{Tsirelson93} (and used in, e.g., Ref.~\cite{BCPSW14}), we will refer to $\mathbf{p}(S)$ as a ``behavior'' for~$S$ and to $\cal {P}(S)$ as the set of behaviors for~$S$. In the literature $\mathbf{p}(S)$ is also called an ``empirical model'' \cite{AB11} or a ``probability model'' \cite{AH12,AFLS15}.

For a given Bell or KS scenario $S$, every initial state and set of measurements produce one behavior.
For example, for the CHSH Bell scenario, if we denote the probabilities $P(x=a,y=b|\psi)$ as $P(ab|xy)$, a behavior can be represented by the following matrix:
\begin{equation}
\mathbf{p}(S_{\rm CHSH}) =
\begin{bmatrix}
P(00|00) & P(01|00) & P(10|00) & P(11|00) \\
P(00|01) & P(01|01) & P(10|01) & P(11|01) \\
P(00|10) & P(01|10) & P(10|10) & P(11|10) \\
P(00|11) & P(01|11) & P(10|11) & P(11|11)
\end{bmatrix},
\end{equation}
where each row contains the probabilities of the events of a maximal context.

%%%%%%%%%%%%%%%%%%%%%%%%%%%%%%%%%%%%%%%%%%%%%%%%%%%%%%%%%%%%%%%%%%%

\subsection{Constraints}
\label{constraints}

%%%%%%%%%%%%%%%%%%%%%%%%%%%%%%%%%%%%%%%%%%%%%%%%%%%%%%%%%%%%%%%%%%%

The behaviors for a given Bell or KS scenario $S$ must satisfy three constraints.

(A) Normalization: for every context $\{x,\ldots,z\}$ (with respective outcomes $a \in A,\ldots,c \in C$) in $S$, 
\begin{equation}
\sum_{a \in A,\ldots, c \in C} P(x=a,\ldots,z=c|\psi)=1.
\end{equation}

(B) Nondisturbance: every pair $(X,Y)$ of mutually nondisturbing sets of measurements in~$S$ must satisfy conditions~(\ref{no-disturbance1}) and (\ref{no-disturbance2}). 

(C) The probability of each event of~$S$ must only be a function of the state and measurement outcomes that define this event. For example, $P(x=a,y=b|\psi)$ must only be a function of $\psi$, $x=a$, and $y=b$. 

%%%%%%%%%%%%%%%%%%%%%%%%%%%%%%%%%%%%%%%%%%%%%%%%%%%%%%%%%%%%%%%%%%%

\subsection{Quantum correlations}
\label{qc}

%%%%%%%%%%%%%%%%%%%%%%%%%%%%%%%%%%%%%%%%%%%%%%%%%%%%%%%%%%%%%%%%%%%

According to QT, the only possible behaviors for a Bell or KS scenario $S$ are those that admit a mathematical representation given by the following conditions.

(I) The initial state $\psi$ of the system can be associated with a vector with unit norm $| \psi \rangle$ in a vector space ${\mathcal V}$ with an inner product.

(II) The state of the system after performing a set $\{x^{(i)}\}$ of compatible measurements in $S$ with respective outcomes $\{a_i\}$ can be associated with a vector in ${\mathcal V}$ with unit norm 
\begin{equation}
\label{2.6}
|\psi'\rangle = N_i \prod_i E^{(i)}_{a_i} |\psi\rangle,
\end{equation}
where $N_i$ is a normalization constant and $\{E^{(i)}_{a_i}\}$ are projection operators. $E^{(i)}_{a_j}$ corresponds to measurement $x^{(i)}$ in $S$ with outcome~$a_j$. Different outcomes of the same measurement must be represented by orthogonal projectors. That is,
\begin{equation}
E^{(i)}_{a_j} E^{(i)}_{a_k} = \delta_{j,k} E^{(i)}_{a_k}.
\end{equation}
The sum of the projectors corresponding to all outcomes is the identity, i.e.,
\begin{equation}
\sum_k E^{(i)}_{a_k}=\id.
\end{equation}
If two measurements $x^{(i)}$ and $x^{(k)}$ are compatible, then 
\begin{equation}
[E^{(i)}_{a_j},E^{(k)}_{a_m}]=0\; \forall j,m,
\end{equation}
where $[\ldots]$ denotes the commutator.

(III) The probability of obtaining $\{a_i\}$ when measuring $\{x^{(i)}\}$ on state $\psi$ satisfies $| \langle \psi' | \psi \rangle |^2$, where $|\psi' \rangle$ is given by Eq.~(\ref{2.6}).

The mathematical characterization of the quantum behaviors is identical for Bell scenarios (in which we do not assume that measurements are ideal) and for KS scenarios (in which all measurements are ideal by definition). This reflects the fact that any quantum behavior for a Bell scenario can be attained with ideal measurements. This follows from Neumark's dilation theorem \cite{Neumark40,Holevo80,Peres95} that shows that every generalized measurement in QT [represented by a positive-operator valued measure (POVM)] can be implemented as an ideal quantum measurement [represented by a projection-valued measure (PVM)] on a larger Hilbert space. In a Bell scenario, any local POVM $x$ admits a local dilation to a PVM that is common to every context in which $x$ appears.

The mathematical characterization of quantum correlations for Bell and KS scenarios given by~(I)--(III) provides no clue of what is the physical reason that selects some behaviors and forbids others. The problem of the physical origin of quantum correlations is precisely identifying what leads to~(I)--(III).

%%%%%%%%%%%%%%%%%%%%%%%%%%%%%%%%%%%%%%%%%%%%%%%%%%%%%%%%%%%%%%%%%%%

\subsection{Independent experiments}
\label{ie}

%%%%%%%%%%%%%%%%%%%%%%%%%%%%%%%%%%%%%%%%%%%%%%%%%%%%%%%%%%%%%%%%%%%

{\em Definition 7.}
	Two experiments ${\cal A}$ and ${\cal B}$ are statistically independent if the occurrence of any of the events of ${\cal A}$ (${\cal B}$) does not affect the probability of occurrence of any of the events of ${\cal B}$ (respectively, ${\cal A}$).

Therefore, if experiment ${\cal A}$ occurs in scenario $S_A$ and produces a behavior $\mathbf{p}(S_A)$, and a statistically independent experiment ${\cal B}$ occurs in scenario $S_B$ and produces a behavior $\mathbf{p'}(S_B)$, then an observer can define an experiment $({\cal A},{\cal B})$ producing a behavior given by the matrix 
\begin{equation}
\label{iex}
\mathbf{p}''(S_{(A,B)}) = \mathbf{p}(S_A) \otimes \mathbf{p'}(S_B),
\end{equation}
where $\otimes$ denotes tensor product. Each row in $\mathbf{p}''(S_{(A,B)})$ contains the probabilities of the events of a maximal context.

If $G_A$ and $G_B$ are the graphs of compatibility of scenarios $S_A$ and $S_B$, respectively, then the graph of compatibility of scenario $S_{(A,B)}$ is the graph join of $G_A$ and $G_B$, denoted $G_A \triangledown G_B$, that is, the graph with vertex set $V(G_A \triangledown G_B) = V(G_A) \cup V(G_B)$ and edge set $E(G_A \triangledown G_B) = E(G_A) \cup E(G_B) \cup \{(v_A,v_B) : v_A \in V(G_A), v_B \in V(G_B)\}$.

Two events $(i, i')$ and $(j, j')$ are mutually exclusive in $S_{(A,B)}$ if $i$ and $j$ are mutually exclusive in $S_A$ or $i'$ and $j'$ are mutually exclusive in $S_B$. Therefore, if $G_{S_A}$ and $G_{S_B}$ are the graphs of exclusivity of the events of $S_A$ and $S_B$, respectively, then the graph of exclusivity of the events of $S_{(A,B)}$ is the OR product of $G_{S_A}$ and $G_{S_B}$, denoted $G_{S_A} \ast G_{S_B}$, that is, the graph with vertex set $V(G_{S_A} \ast G_{S_B}) = V(G_{S_A}) \times V(G_{S_B})$ and edge set $E(G_{S_A} \ast G_{S_B}) = \{((i, i'), (j, j')) : (i,j) \in E(G_{S_A})\;{\rm or}\;(i',j') \in E(G_{S_B})\}$.

%%%%%%%%%%%%%%%%%%%%%%%%%%%%%%%%%%%%%%%%%%%%%%%%%%%%%%%%%%%%%%%%%%%

\section{Result}
\label{Sec3}

%%%%%%%%%%%%%%%%%%%%%%%%%%%%%%%%%%%%%%%%%%%%%%%%%%%%%%%%%%%%%%%%%%%

{\em Assumption 1.}
There is a nonempty set of behaviors for any KS scenario.

{\em Assumption 2.}
There is a statistically independent joint realization of any two KS~experiments.

{\em Theorem 1.}
The set of behaviors ${\cal P}(S)$ allowed by quantum theory for any Bell or KS scenario~$S$ is equal to the largest set allowed by Assumptions~1 and~2.

Assumption~1 also applies to Bell scenarios with ideal measurements. Assumption~2 also applies to Bell experiments with arbitrary measurements.

Assumption~1 emphasizes ideal measurements. However, it does not require that ideal measurements are physically achievable. Even though actual measurement processes would fail to exactly satisfy conditions (i), (ii), and (iii) in Definition~4, Assumption~1 only requires the {\em theory} to assign probabilities to the outcomes of these idealized measurement processes. Arguably, Assumption~1 is inescapable in any theory that contains classical physics as a particular case. This is so because, in classical physics, a measurement is an interaction between a physical system and a measuring device that reveals preexisting, persistent, context-independent properties of the physical system and that allows subsequent interactions revealing fine-grained (or coarse-grained) preexisting, persistent, context-independent properties of the same physical system.

Assumption~2 emphasizes statistical independence. However, it does not preclude the existence of a universe massively interconnected through causal chains and full of strongly correlated experiments. It only assumes that, even within such a universe, there is a statistically independent joint realization of any two KS~experiments. Arguably, this form of statistical independence holds in every physical theory.

%%%%%%%%%%%%%%%%%%%%%%%%%%%%%%%%%%%%%%%%%%%%%%%%%%%%%%%%%%%%%%%%%%%

\section{Proof}
\label{Sec4}

%%%%%%%%%%%%%%%%%%%%%%%%%%%%%%%%%%%%%%%%%%%%%%%%%%%%%%%%%%%%%%%%%%%

\subsection{Exclusivity principle}

%%%%%%%%%%%%%%%%%%%%%%%%%%%%%%%%%%%%%%%%%%%%%%%%%%%%%%%%%%%%%%%%%%%

{\em Lemma 1.}
	The behaviors for any KS scenario must satisfy the exclusivity principle~(EP). 

{\em Lemma 2.}
	The behaviors for any bipartite Bell scenario must satisfy the EP.

A behavior for scenario $S$ satisfies the EP \cite{CSW10,Cabello13,Yan13,CSW14,ATC14,CY14,Cabello15,Henson15,AFLS15} if, for every subset of events $\{e_i\}$ in $S$ such that every two events in $\{e_i\}$ are mutually exclusive in $S$, their probabilities satisfy
\begin{equation}
\sum_{i} P(e_i) \leq 1.
\end{equation}
The EP is also called the principle of global exclusivity \cite{Cabello13,Yan13} or consistent exclusivity \cite{Henson15,AFLS15}. 

The EP does not follow from the axioms of probability [namely: (a)~the probability of an event is a non-negative real number, (b)~the probability that at least one of the elementary events in the entire sample space will occur is $1$, and (c)~the probability of the union of any countable sequence of mutually exclusive events is the sum of their probabilities]. Notice the difference between a set of mutually exclusive events $\{e_1,e_2,\ldots,e_n\}$ and a set of events $\{e'_1,e'_2,\ldots,e'_n\}$ in which every two events are mutually exclusive. In the first case, there is a single measure $M$ that produces $e_1$, $e_2,\ldots,e_n$, each of them associated to a different outcome of $M$. In the second case, there are $\binom{n}{2}$ different measurements $M_{ij}$, with $(i,j) \in \{(1,2),(1,3),\ldots,(1,n),(2,3),\ldots,(2,n),\ldots,(n-1,n)\}$, such that $M_{ij}$ produces $e'_i$ and $e'_j$, each of them associated to a different outcome of $M_{ij}$.

It is important to stress that, although we will continue to refer to the EP as a ``principle,'' in the proof of Theorem~1 we do not {\em assume} the EP. Instead, we will invoke Lemmas~1 and~2. Lemma~1 follows from a result in Ref.~\cite{CY14}. Other proof of this result can be found in Ref.~\cite{CCKM19}. For completeness' sake, we include a proof of Lemma~1 in Appendix~\ref{AppB}. Lemma~2 was proven in Ref.~\cite{CSW10}. Other proof can be found in Ref.~\cite{FSABCLA13}. For completeness' sake, we include a proof of Lemma~2 in Appendix~\ref{AppA}.

%%%%%%%%%%%%%%%%%%%%%%%%%%%%%%%%%%%%%%%%%%%%%%%%%%%%%%%%%%%%%%%%%%%

\subsection{Combining Assumption~2 and Lemmas~1 and~2}

%%%%%%%%%%%%%%%%%%%%%%%%%%%%%%%%%%%%%%%%%%%%%%%%%%%%%%%%%%%%%%%%%%%

%%%%%%%%%%%%%%%%%%%%%%%%%%%%%%%%%%%%%%%%%%%%%%%%%%%%%%%%%%%%%%%%%%%
% Fig. 2
%%%%%%%%%%%%%%%%%%%%%%%%%%%%%%%%%%%%%%%%%%%%%%%%%%%%%%%%%%%%%%%%%%%

\begin{figure*}[t]
	%\vspace{0mm}
	\hspace{-4.4mm}
	\includegraphics[width=16.8cm]{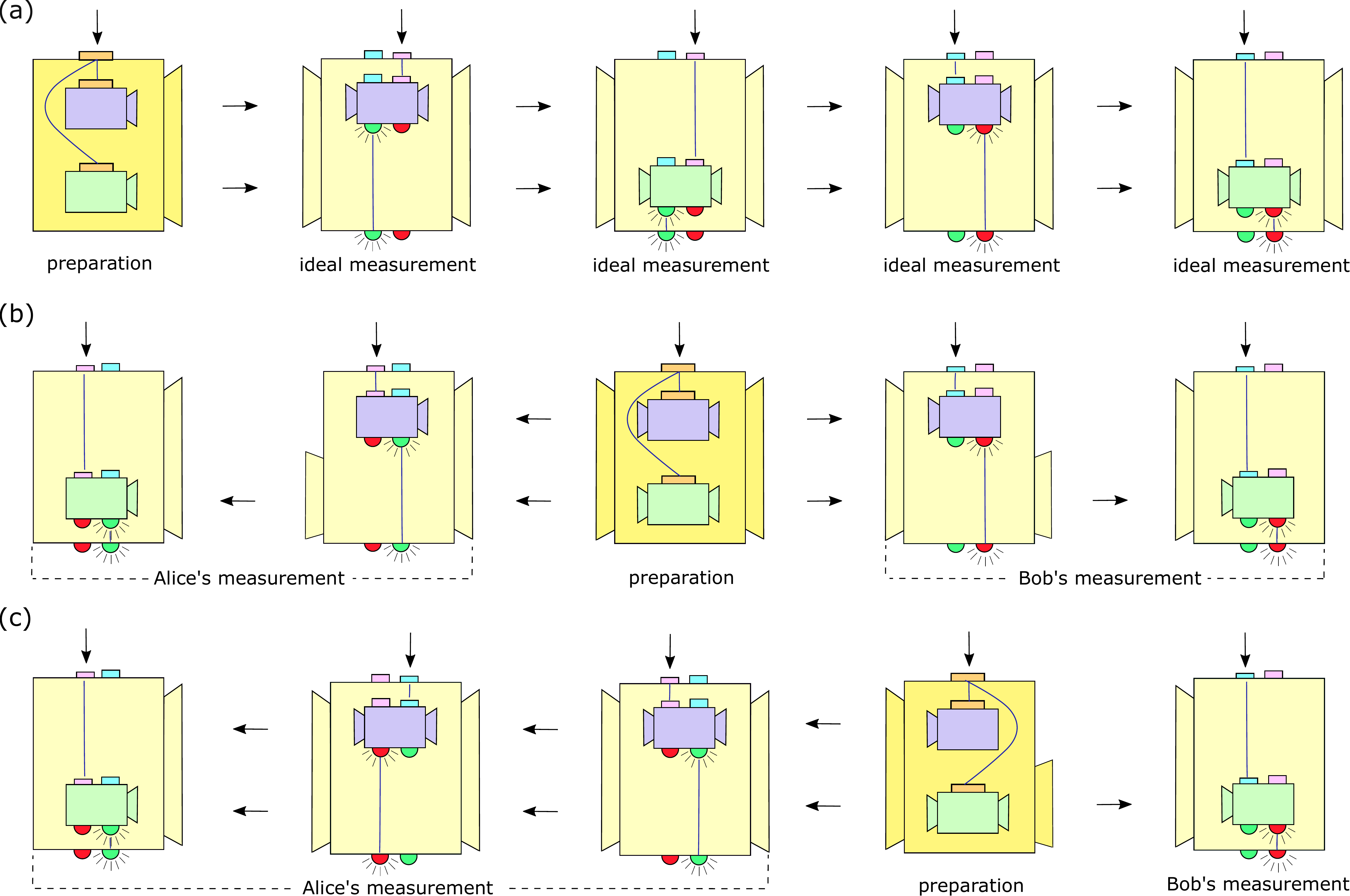}
	\caption{(a)~Experiment $({\cal A},{\cal B})$ defined from the joint realization of two statistically independent KS experiments ${\cal A}$ and ${\cal B}$ seen as a single KS experiment. The purple boxes represent experiment ${\cal A}$ and the green boxes experiment ${\cal B}$. Experiments in different horizontal lines are statistically independent. The yellow boxes represent $({\cal A},{\cal B})$. Pressing the button at the top of a preparation box releases a physical system. Pressing a button at the top of a measurement box selects a measurement. The outcome of this measurement is represented by the bulb that flashes at the bottom. Pressing a button of a yellow box corresponds to pressing a button of the small box that is inside it. The outcome of the yellow box corresponds to an outcome of the small box inside it. (b)~Experiment $({\cal A},{\cal B})$ defined from the joint realization of two statistically independent bipartite Bell experiments ${\cal A}$ (in purple) and ${\cal B}$ (in green) seen as a single bipartite Bell experiment (in yellow). The meaning of the symbols is the same as in case~(a). Here, measurements are not necessarily ideal. (c)~Experiment $({\cal A},{\cal B})$ defined from the joint realization of a KS experiment ${\cal A}$ (in purple) and a bipartite Bell experiment ${\cal B}$ (in green) seen as a single bipartite Bell experiment (in yellow).}
	\label{Fig2}
\end{figure*}

%%%%%%%%%%%%%%%%%%%%%%%%%%%%%%%%%%%%%%%%%%%%%%%%%%%%%%%%%%%%%%%%%%%

Assumption~2 assures that there is a statistically independent joint realization of any two KS experiments ${\cal A}$~and~${\cal B}$ (including Bell experiments). This allows us to define experiments of the type $({\cal A},{\cal B})$ described in Sec.~\ref{ie}.

If ${\cal A}$ and ${\cal B}$ are KS experiments, then $({\cal A},{\cal B})$ can be seen as a single KS experiment. See Fig.~\ref{Fig2}(a). Therefore, Lemma~1 assures that the behaviors for $({\cal A},{\cal B})$ must satisfy the~EP. 

If ${\cal A}$ and ${\cal B}$ are bipartite Bell experiments, then $({\cal A},{\cal B})$ can be seen as a bipartite Bell experiment. See Fig.~\ref{Fig2}(b). Therefore, Lemma~2 assures that the behaviors for $({\cal A},{\cal B})$ must satisfy the~EP.

If ${\cal A}$ is a KS experiment and ${\cal B}$ a bipartite Bell experiment, then $({\cal A},{\cal B})$ can be seen as a bipartite Bell experiment. See Fig.~\ref{Fig2}(c). Therefore, Lemma~2 assures that the behaviors for $({\cal A},{\cal B})$ must satisfy the~EP.

The same reasoning can be applied to $n$~statistically independent experiments. A particularly interesting case is when the same behavior $\mathbf{p}(S_A)$ for a Bell or KS~experiment is composed with itself $n$~times. If $\mathbf{p}(S_A)$ occurs in $n$~statistically independent experiments ${\cal A}_i$, then, by Lemmas~1 and~2, the behavior for the corresponding experiment $({\cal A}_1, \ldots, {\cal A}_n)$, that is,
\begin{equation}
\mathbf{p}(S_{(A_1,\ldots,A_n)})=\mathbf{p}(S_{\cal A})^{\otimes n},
\end{equation}
where $\mathbf{p}(S_{\cal A})^{\otimes n}$ denotes the tensor product of $n$ copies of $\mathbf{p}(S_{\cal A})$, must satisfy the EP for any $n \in \mathbb{N}$.

%%%%%%%%%%%%%%%%%%%%%%%%%%%%%%%%%%%%%%%%%%%%%%%%%%%%%%%%%%%%%%%%%%% 

\subsection{Assignments of probabilities for a graph of exclusivity}
\label{cond}

%%%%%%%%%%%%%%%%%%%%%%%%%%%%%%%%%%%%%%%%%%%%%%%%%%%%%%%%%%%%%%%%%%%

Here is where the unified treatment of different KS and Bell scenarios allowed by the graph-theoretic approach \cite{CSW10,CSW14} enters into the proof of Theorem~1. Given any graph $G$ with vertex set $V(G)$, we can consider the following set.

{\em Definition 8.}
	The set ${\cal P}(G)$ of assignments of probabilities to the vertices of graph~$G$ is the set of vectors $\mathbf{p}(G) \in [0,1]^{|V(G)|}$ such that the components of $\mathbf{p}(G)$ are the probabilities of $|V(G)|$ events with graph of exclusivity~$G$ in a behavior for {\em some} Bell or KS scenario.

That is, ${\cal P}(G)$ contains the vectors of probabilities with $|V(G)|$ components corresponding to events that have~$G$ as graph of exclusivity produced in {\em all} Bell and KS scenarios. Hereafter, we will refer to $\mathbf{p}(G)$ as an assignment of probabilities to the vertices of $G$ (or, for brevity, an assignment of probabilities for $G$).

As shown before, the behaviors for KS scenarios must satisfy the EP. Therefore, for any $G$, ${\cal P}(G)$ must be a subset of
\begin{equation}
\qstab(G) = \{ \mathbf{p}(G) \in [0,1]^{|V(G)|} : \sum_{i\in c} p_i \le 1\,\;\forall c \in C(G) \},
\end{equation}
where $C(G)$ is the set of cliques of $G$ \cite{CSW14}. $\qstab(G)$ is a famous set in graph theory called the fractional vertex packing polytope \cite{GLS86}, or clique-constrained stable set polytope \cite{GLS88}, or fractional stable set polytope \cite{GLS88} of $G$.

For any theory, ${\cal P}(G)$ for each $G$ captures a fundamental signature of the theory. 
For example, in QT, ${\cal P}(G)$ is the set of assignments that satisfy the following.

(I') The initial state of the system can be associated with a vector with unit norm $| \psi \rangle$ in a vector space ${\mathcal V}$ with an inner product.

(II') The state of the system after performing a measurement $x^{(i)}$ and obtaining outcome $a_i$ on state $\psi$ can be associated with a vector with unit norm $ |x^{(i)}_{a_i}\psi \rangle$ in ${\mathcal V}$. Postmeasurement states corresponding to mutually exclusive events are associated with mutually orthogonal vectors.

(III') The probability of event $(x^{(i)}_{a_i} | \psi)$ can be obtained as $|\langle x^{(i)}_{a_i} \psi | \psi \rangle|^2$.

Surprisingly, this set is a well-known convex set in graph theory \cite{CSW14}: the theta body of $G$, denoted $\thb(G)$ \cite{GLS86,GLS88,Knuth94}. Using Dirac's notation,
\begin{equation}
\label{thb}
\begin{split}
\thb(G) =& \{ \mathbf{p}(G) \in [0,1]^{|V(G)|} : p_i = |\langle x^{(i)}_{a_i} \psi | \psi \rangle|^2,\; \\
&
|\langle \psi| \psi \rangle| = 1, |\langle x^{(i)}_{a_i}\psi | x^{(i)}_{a_i}\psi \rangle| = 1, \\
&
\langle x^{(i)}_{a_i}\psi | x^{(j)}_{a_j}\psi \rangle = 0\;\forall (i,j) \in E(G) \},
\end{split}
\end{equation}
where $E(G)$ is the set of edges of $G$.

A famous result in graph theory is that $\thb(G)=\qstab(G)$ if and only if $G$ does not contain odd cycles with five or more vertices (i.e., pentagons, heptagons, etc.) or their complements as induced subgraphs. These graphs are called perfect graphs \cite{GLS88,Knuth94}. This means that the EP, by itself, selects the quantum set of assignments for perfect graphs. Recall that, given a graph $G$, the complement of $G$, denoted $\overline{G}$, is the graph with the same vertices as $G$ such that two distinct vertices of $\overline{G}$ are adjacent if and only if they are not adjacent in $G$. 

A fundamental problem proposed in Ref.~\cite{CSW14} is identifying the principle that selects $\thb(G)$ for arbitrary graphs. 
The next step in the proof of Theorem~1 is precisely showing that Assumptions~1 and~2 select $\thb(G)$ for any $G$.

%%%%%%%%%%%%%%%%%%%%%%%%%%%%%%%%%%%%%%%%%%%%%%%%%%%%%%%%%%%%%%%%%%%

\subsection{EP applied to independent copies of an assignment}

%%%%%%%%%%%%%%%%%%%%%%%%%%%%%%%%%%%%%%%%%%%%%%%%%%%%%%%%%%%%%%%%%%%

Assignments of probabilities that might look possible at a first sight can be excluded by Assumption~2 and Lemmas~1 and~2. Here we illustrate this with an example. We will consider an increasing number of statistically independent copies of an assignment $\mathbf{p}(G)$ and identify an increasing set of assignments that are incompatible with Assumptions~1 and~2. 

%%%%%%%%%%%%%%%%%%%%%%%%%%%%%%%%%%%%%%%%%%%%%%%%%%%%%%%%%%%%%%%%%%%
% Fig. 3
%%%%%%%%%%%%%%%%%%%%%%%%%%%%%%%%%%%%%%%%%%%%%%%%%%%%%%%%%%%%%%%%%%%

\begin{figure*}[t!]
	\vspace{-27.6mm}
	\hspace{-26.0mm}
	\includegraphics[width=20cm]{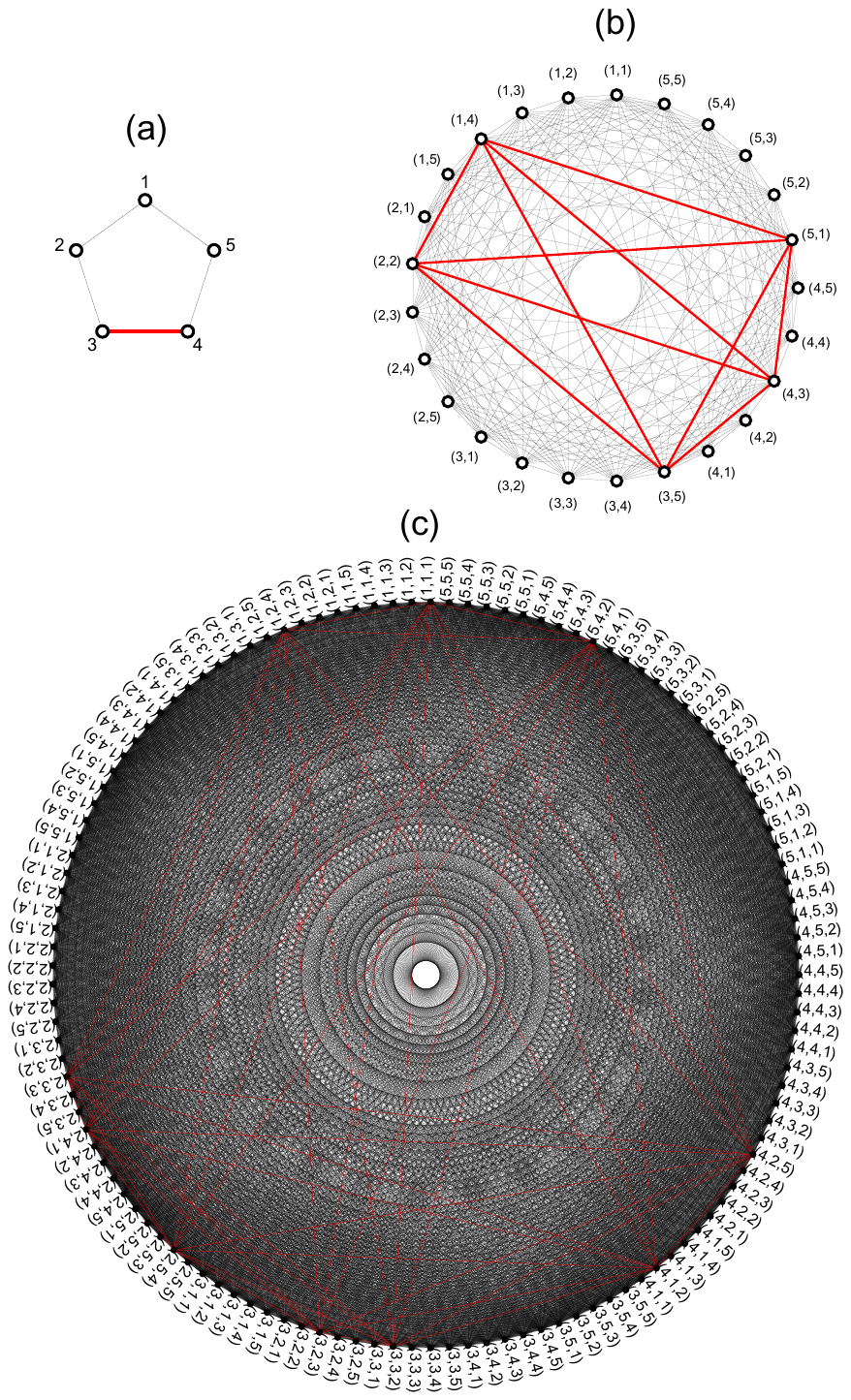}
	\vspace{-23.0mm}
	\caption{(a)~Graph of exclusivity of five events of experiment~${\cal A}$. (b)~Graph of exclusivity of the corresponding $25$~events of experiment~$({\cal A},{\cal A})$. (c)~Graph of exclusivity of the corresponding $125$~events of experiment~$({\cal A},{\cal A},{\cal A})$. The red edges join the events that are mutually exclusive pairwise that are used for the arguments in the text.}
	\label{Fig3}
\end{figure*}

%%%%%%%%%%%%%%%%%%%%%%%%%%%%%%%%%%%%%%%%%%%%%%%%%%%%%%%%%%%%%%%%%%%

Consider the following set of assignments of probabilities to the events of an experiment ${\cal A}$ whose graph of exclusivity is the pentagon (denoted $C_5$) shown in Fig.~\ref{Fig3}(a):
\begin{equation}
\label{Penta2}
\mathbf{p} (C_5)= \left \{ \tfrac{1}{3},\tfrac{2}{3},x,\tfrac{2}{3},\tfrac{1}{3} \right \},
\end{equation}
where the components are ordered following the labeling of the vertices in Fig.~\ref{Fig3}(a).
If we apply the EP to the clique $\{3,4\}$ in red in Fig.~\ref{Fig3}(a), then the EP enforces that 
\begin{equation}
x+\tfrac{2}{3} \leq 1.
\end{equation}
Therefore, the EP excludes any assignment with
\begin{equation}
x > \tfrac{1}{3} \approx 0.333.
\end{equation}
That is, none of these assignments can be produced under Assumptions~1 and~2.

Now let's consider the joint realization of two statistically independent copies of $\mathbf{p}(C_5)$ as seen from the observer contemplating experiment $({\cal A},{\cal A})$.
Recall that the graph of exclusivity corresponding to two copies of an experiment whose graph of exclusivity is $G$ is given by the OR product $G \ast G$. In Fig.~\ref{Fig3}(b) we show the OR product of two copies of the pentagon in Fig.~\ref{Fig3}(a). 

If we apply the EP to the clique $\{(1,4),(2,2),(3,5),(4,3),(5,1)\}$ in red in Fig.~\ref{Fig3}(b), then the EP enforces that
\begin{equation}
x+\tfrac{7}{9} \leq 1.
\end{equation}
Therefore, the EP excludes any assignment with
\begin{equation}
x > \tfrac{2}{9} \approx 0.222.
\end{equation}
No other clique of the graph in Fig.~\ref{Fig3}(b) allows us to further restrict the values of $x$ using the~EP.

Now lets consider the joint realization of three statistically independent copies of $\mathbf{p}(C_5)$ as seen from the observer contemplating experiment $({\cal A},{\cal A},{\cal A})$. The graph of exclusivity corresponding to three copies of an experiment whose graph of exclusivity is $G$ is given by $G \ast G \ast G$. In Fig.~\ref{Fig3}(c) we show the OR product of three copies of the pentagon in Fig.~\ref{Fig3}(a). 

If we apply the EP to the clique $\{(1, 1, 1)$, $(1, 2, 4)$, $(2, 3, 3)$, $(2, 4, 1)$, $(2, 5, 4)$, $(3, 2, 3)$, $(3, 3, 2)$, $(4, 1, 2)$, $(4, 2, 5)$, $(5, 4, 2)\}$ in red in Fig.~\ref{Fig3}(c), then the EP enforces that
\begin{equation}
2 x^2+\tfrac{25}{27} \leq 1.
\end{equation}
Therefore, the EP excludes any assignment with
\begin{equation}
x > \tfrac{1}{3 \sqrt{3}}\approx 0.192.
\end{equation}
No other clique of the graph in Fig.~\ref{Fig3}(c) allows us to further restrict the values of $x$ using the~EP.

The EP excludes more assignments as we consider more independent copies of $\mathbf{p}(G)$. The problem is that finding which assignments are excluded is increasingly hard as we consider more copies. In fact, finding the largest clique of $G^{\ast n}$, for $n \ge 4$ is a very difficult problem even for very small graphs (see, e.g., Ref.~\cite{BH03}).
Interestingly, for some special graphs we can characterize the set of assignments that survive after we apply the EP to any number of independent copies. 

%%%%%%%%%%%%%%%%%%%%%%%%%%%%%%%%%%%%%%%%%%%%%%%%%%%%%%%%%%%%%%%%%%%

\subsection{Characterizing the quantum assignments for self-complementary graphs of exclusivity}

%%%%%%%%%%%%%%%%%%%%%%%%%%%%%%%%%%%%%%%%%%%%%%%%%%%%%%%%%%%%%%%%%%%

A graph $G$ is self-complementary if $G$ and its complement $\overline{G}$
are isomorphic. 
In this case we will write that $G=\overline{G}$. 
For example, the pentagon and its complement, the pentagram, are isomorphic; therefore, the pentagon is self-complementary.

The fact that any assignment $\mathbf{p}(G)$ must satisfy the EP applied to 
any number $n$ of independent copies of $\mathbf{p}(G)$ implies
that $\mathbf{p}(G)^{\,\otimes n}$ must satisfy the EP for any $n \in \mathbb{N}$.
Therefore, for any $n \in \mathbb{N}$, the set of assignments for $G$ must satisfy
\begin{equation}
\label{Cond1}
{\cal P}(G) \subseteq {\cal E}^n(G),
\end{equation}
where
\begin{equation}
{\cal E}^n(G) = \{\mathbf{p}(G) \in [0,1]^{|V(G)|} : \mathbf{p}(G)^{\,\otimes n} \in \qstab(G^{\ast n})\}.
\end{equation}

On the other hand, any $\mathbf{p}(G), \mathbf{p}'(G) \in {\cal P}(G)$ must satisfy the EP applied to one copy of $\mathbf{p}(G)$ and one independent copy of $\mathbf{p}'(G)$. This implies that
\begin{equation}
\label{Cond2}
{\cal P}(G) \subseteq \abl[{\cal P}(G)],
\end{equation} 
where $\abl[{\cal P}(G)]$ is the antiblocker of ${\cal P}(G)$, defined as 
\begin{equation}
\abl[{\cal P}(G)]= \{\mathbf{p}'(G) \geq \mathbf{0} : \mathbf{p}(G) \cdot \mathbf{p}'(G) \leq 1\,\forall \mathbf{p}(G) \in {\cal P}(G) \},
\end{equation}
where $\cdot$ is the dot product \cite{GLS88,Knuth94,GLS86}.

{\em Lemma~3.}
For any self-complementary graph of exclusivity $G$, the theta body of $G$, $\thb(G)$, is the largest set of assignments of probabilities ${\cal P}(G)$ such that every $\mathbf{p}(G) \in {\cal P}(G)$ satisfies the EP applied to any number of independent copies of $\mathbf{p}(G)$ and such that $\mathbf{p}(G) \otimes \mathbf{p}'(G)$ satisfies the EP for every $\mathbf{p}(G), \mathbf{p}'(G) \in {\cal P}(G)$.

\begin{proof}
	If $G=\overline{G}$, then, for any $n \in \mathbb{N}$ \cite{GLS88,Knuth94}, 
	\begin{equation}
	\label{Conv1}
	\abl[{\cal E}^{n-1}(G)] \subseteq \abl[{\cal E}^{n}(G)] \subseteq {\cal E}^{n}(G) \subseteq {\cal E}^{n-1}(G).
	\end{equation} 
	Therefore, Eq.~(\ref{Cond1}) implies that
	\begin{equation}
	\label{Conv2}
	{\cal P}(G) \subseteq \lim_{n \to \infty } {\cal E}^n(G).
	\end{equation}
	and (\ref{Cond2}) implies that
	\begin{equation}
	\label{Conv3}
	{\cal P}(G) \subseteq \abl[\lim_{n \to \infty } {\cal E}^n(G)].
	\end{equation}
	All sets satisfying (\ref{Conv2}) and (\ref{Conv3}) are subsets of $\thb(G)$. Notice that, if $G=\overline{G}$, then $\thb(G) = \abl[\thb(G)]$ \cite{GLS86,GLS88,Knuth94}.
\end{proof}

Lemma~3 is a very interesting result as it explains why vector spaces and the Born rule appear, and is capable of identifying the origin of some quantum sets of assignments. The problem is that Lemma~3 only covers the case of graphs of exclusivity that are self-complementary.

%%%%%%%%%%%%%%%%%%%%%%%%%%%%%%%%%%%%%%%%%%%%%%%%%%%%%%%%%%%%%%%%%%%

\subsection{Characterizing the quantum assignments for arbitrary graphs of exclusivity}

%%%%%%%%%%%%%%%%%%%%%%%%%%%%%%%%%%%%%%%%%%%%%%%%%%%%%%%%%%%%%%%%%%%

Our next result solves that problem.

{\em Lemma~4.}
	For any graph of exclusivity $G$, $\thb(G)$ is the largest set ${\cal P}(G)$ of assignments of probabilities such that every $\mathbf{p}(G) \in {\cal P}(G)$ satisfies the EP applied to any number of independent copies of $\mathbf{p}(G)$ and such that $\mathbf{p}(G) \otimes \mathbf{p}'(G)$ satisfies the EP for every $\mathbf{p}(G), \mathbf{p}'(G) \in {\cal P}(G)$.

%%%%%%%%%%%%%%%%%%%%%%%%%%%%%%%%%%%%%%%%%%%%%%%%%%%%%%%%%%%%%%%%%%%
% Fig. 4
%%%%%%%%%%%%%%%%%%%%%%%%%%%%%%%%%%%%%%%%%%%%%%%%%%%%%%%%%%%%%%%%%%%

\begin{figure}[t!]
	\hspace{-1mm}
	\includegraphics[width=8.7cm]{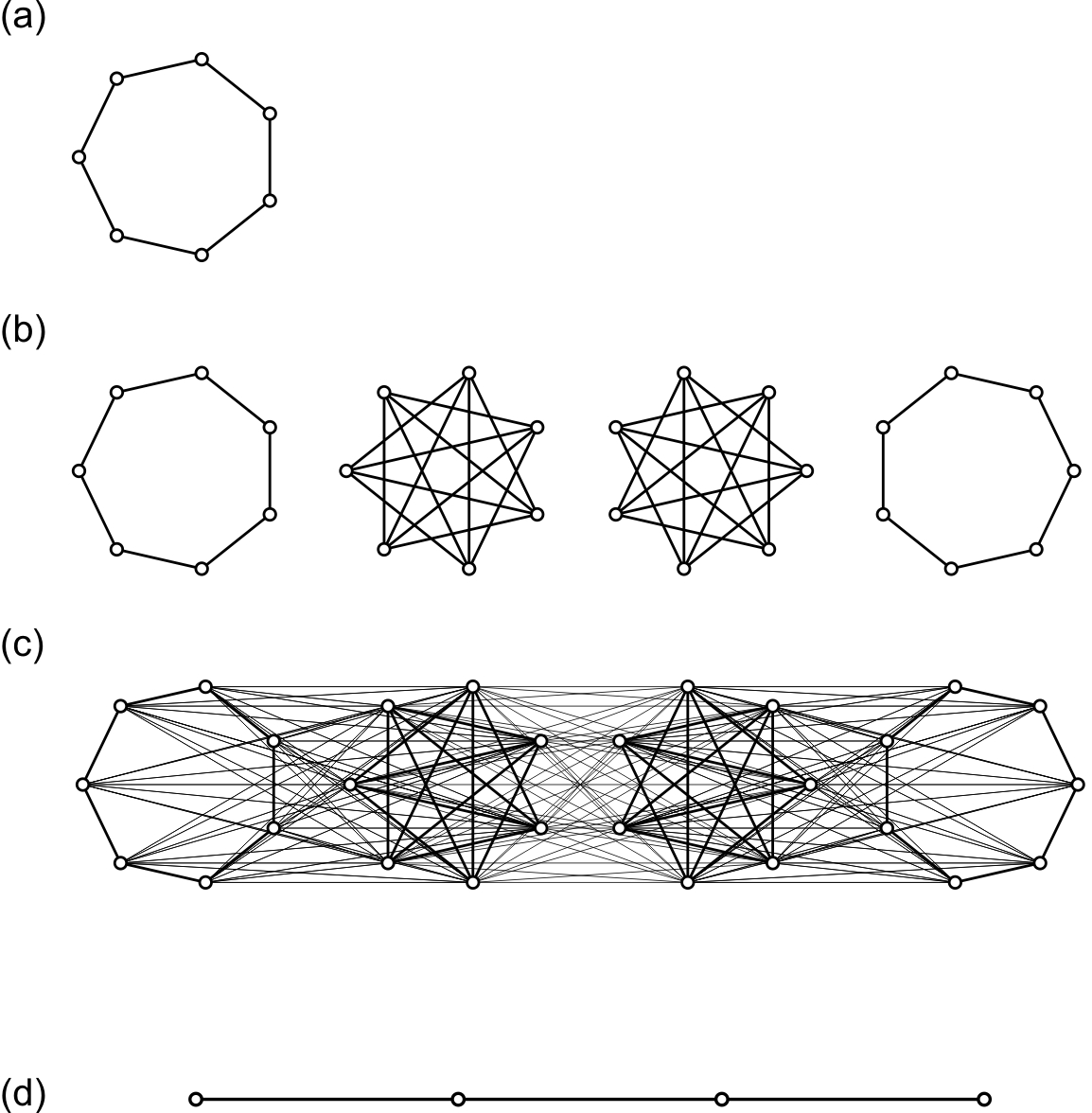}
	\vspace{3mm}
	\caption{Details of the construction of graph $H(G)$ used in the proof of Lemma~4. (a)~Graph of exclusivity $G$ whose set of assignments we are interested in. (b)~Graphs of exclusivity of the events of the experiments ${\cal E}$, ${\cal X}$, ${\cal Y}$, and ${\cal Z}$. (c)~$H(G)$. (d)~Graph ${\mathcal G}$ such that $H(G)={\mathcal G}[G,\overline{G},\overline{G},G]$. See details in the text.}
	\label{Fig4}
\end{figure}

%%%%%%%%%%%%%%%%%%%%%%%%%%%%%%%%%%%%%%%%%%%%%%%%%%%%%%%%%%%%%%%%%%%

\begin{proof}
For any graph of exclusivity $G$, there is a larger graph $H(G)$ with the following two properties:

(i) $H(G)$ is self-complementary and therefore Lemma~3 gives the largest ${\cal P}[H(G)]$ allowed by Assumptions~1 and~2. 

(ii) The largest ${\cal P}[H(G)]$ allowed by Assumptions~1 and~2 determines the largest ${\cal P}(G)$ allowed by these assumptions.

Let us see how $H(G)$ is defined. Consider an experiment~${\cal E}$ producing $n$~events $\{e_k\}_{k=1}^n$ whose graph of exclusivity is $G$. See Fig.~\ref{Fig4}(a) for an example. Intentionally, in the example, $G$ (the heptagon) is not perfect and not self-complementary. Then, consider three additional statistically independent experiments: ${\cal X}$, producing events $\{x_k\}_{k=1}^n$ whose graph of exclusivity is $\overline{G}$, ${\cal Y}$, producing events $\{y_k\}_{k=1}^n$ whose graph of exclusivity is $\overline{G}$, and ${\cal Z}$ producing events $\{z_k\}_{k=1}^n$ whose graph of exclusivity is $G$. See Fig.~\ref{Fig4}(b). Suppose an observer contemplating the four experiments and who, in addition, has three independent coins ${\cal A}$, ${\cal B}$, and ${\cal C}$, each of them producing two mutually exclusive events: ${\cal A}$ producing events $a_0$ or $a_1$, ${\cal B}$ producing $b_0$ or $b_1$, and ${\cal C}$ producing $c_0$ or $c_1$. Suppose that this observer uses all these experiments and defines the following $4 n$ events: 
\begin{equation}
\{(a_0,e_k),(a_1,b_0,x_k),(b_1,c_0,y_k),(c_1,z_k)\}_{k=1}^n,
\end{equation}
where, e.g., $(a_0,e_1)$ is the event in which coin ${\cal A}$ gives $a_0$ and experiment ${\cal E}$ gives $e_1$. $H(G)$ is the graph of exclusivity of these $4 n$ events. See Fig.~\ref{Fig4}(c).

$H(G)$ is the generalized composition ${\mathcal G}[G,\overline{G},\overline{G},G]$, where $G$, $\overline{G}$, $\overline{G}$, and $G$ are the graphs in Fig.~\ref{Fig4}(b) and ${\mathcal G}$ is the graph in Fig.~\ref{Fig4}(d) \cite{Schwenk74}. 
If ${\mathcal G}$ is a graph with $n$ vertices, then the graph ${\mathcal G}[G_1,\ldots, G_n]$ is constructed by taking the disjoint graphs $G_1,\ldots, G_n$ and joining every vertex of $G_i$ with every vertex of $G_j$ whenever $v_i$ and $v_j$ are adjacent vertices in ${\mathcal G}$.

%%%%%%%%%%%%%%%%%%%%%%%%%%%%%%%%%%%%%%%%%%%%%%%%%%%%%%%%%%%%%%%%%%%
% Fig. 5
%%%%%%%%%%%%%%%%%%%%%%%%%%%%%%%%%%%%%%%%%%%%%%%%%%%%%%%%%%%%%%%%%%%

\begin{figure}[b]
	%\hspace{-4mm}
	\includegraphics[width=7.5cm]{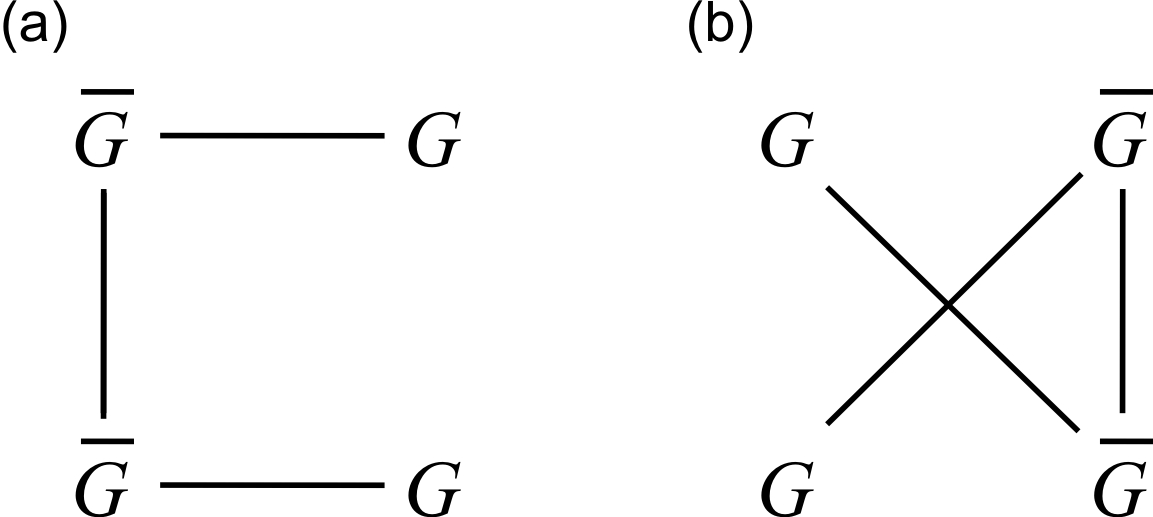}
	\caption{(a) Simplified representation of $H(G)$ as a graph of graphs (i.e., a graph which has a graph in each of its vertices; $G_1-G_2$ denotes that every vertex of $G_1$ is connected to every vertex of $G_2$). (b)~Simplified representation of $\overline{H(G)}$ as a graph of graphs. It is easy to see that $H(G)$ and $\overline{H(G)}$ are isomorphic.}
	\label{Fig5}
\end{figure}

%%%%%%%%%%%%%%%%%%%%%%%%%%%%%%%%%%%%%%%%%%%%%%%%%%%%%%%%%%%%%%%%%%%ç

The first property of $H(G)$, namely that, for any $G$, $H(G)$ is self-complementary, is proven in Fig.~\ref{Fig5}. Then, Lemma~3 gives the largest ${\cal P}[H(G)]$ allowed by Assumptions~1 and~2.

The second property of $H(G)$ comes from the fact that the graph ${\mathcal G}$ in Fig.~\ref{Fig4}(d) is perfect (because it does not have odd cycles of size five or larger or their complements as induced subgraphs). For perfect graphs it has been proven \cite{CSW14} that the set of assignments of probabilities satisfying the EP is equal to the set of assignments for classical (noncontextual) theories. This implies that any assignment $\mathbf{h}[H(G)] \in {\cal P}[H(G)]$ that satisfies the EP can be implemented by suitably choosing an assignment $\mathbf{p}(G)$ for $\{e_k\}_{k=1}^n$, an assignment $\mathbf{x}(\overline{G})$ for $\{x_k\}_{k=1}^n$, an assignment $\mathbf{y}(\overline{G})$ for $\{y_k\}_{k=1}^n$, an assignment $\mathbf{z}(G)$ for $\{z_k\}_{k=1}^n$, an assignment $\mathbf{a}(K_2)$ for $\{a_0,a_1\}$ ($K_2$ is the complete graph on two vertices; the graph of exclusivity of the events of tossing a coin), an assignment $\mathbf{b}(K_2)$ for $\{b_0,b_1\}$, and an assignment $\mathbf{c}(K_2)$ for $\{c_0,c_1\}$. In other words,
\begin{equation}
\label{perfe0}
\begin{split}
& {\cal P}[H(G)] = \text{convex hull} \{\mathbf{h}[H(G)]=(\mathbf{p}(G),\mathbf{x}(G),\mathbf{y}(G),\mathbf{z}(G)) \\
& \in \{({\cal P}(G),0^{|V(G)|},0^{|V(G)|},{\cal P}(G)), \\
& ({\cal P}(G),0^{|V(G)|},{\cal P}(\overline{G}),0^{|V(G)|}), \\
& (0^{|V(G)|},{\cal P}(\overline{G}),0^{|V(G)|},{\cal P}(G)) \}\},
\end{split}
\end{equation}
where ${\cal P}(G)$ is the set of assignments for $G$ such that $\mathbf{p}(G) \in {\cal P}(G)$ satisfies the EP applied to any number of statistically independent copies of $\mathbf{p}(G)$ and $\mathbf{p}(G) \otimes \mathbf{p}'(G)$, with $\mathbf{p}(G), \mathbf{p}'(G) \in {\cal P}(G)$, satisfies the EP.
Notice that the convex hull plays the role of $\mathbf{a}(K_2)$, $\mathbf{b}(K_2)$, and $\mathbf{c}(K_2)$.
Therefore, the largest ${\cal P}(G)$ allowed by Assumptions~1 and~2 can be obtained from the largest ${\cal P}[H(G)]$ allowed by these assumptions by suitably tracing out its elements.
\end{proof}

Therefore, Lemma~4 solves the problem proposed in Ref.~\cite{CSW14}: Assumptions~1 and~2 select the quantum sets of probability assignments for {\em any} graph of exclusivity.

%%%%%%%%%%%%%%%%%%%%%%%%%%%%%%%%%%%%%%%%%%%%%%%%%%%%%%%%%%%%%%%%%%%
% Fig. 6
%%%%%%%%%%%%%%%%%%%%%%%%%%%%%%%%%%%%%%%%%%%%%%%%%%%%%%%%%%%%%%%%%%%

\begin{figure}
	%\hspace{-4mm}
	\includegraphics[width=8.0cm]{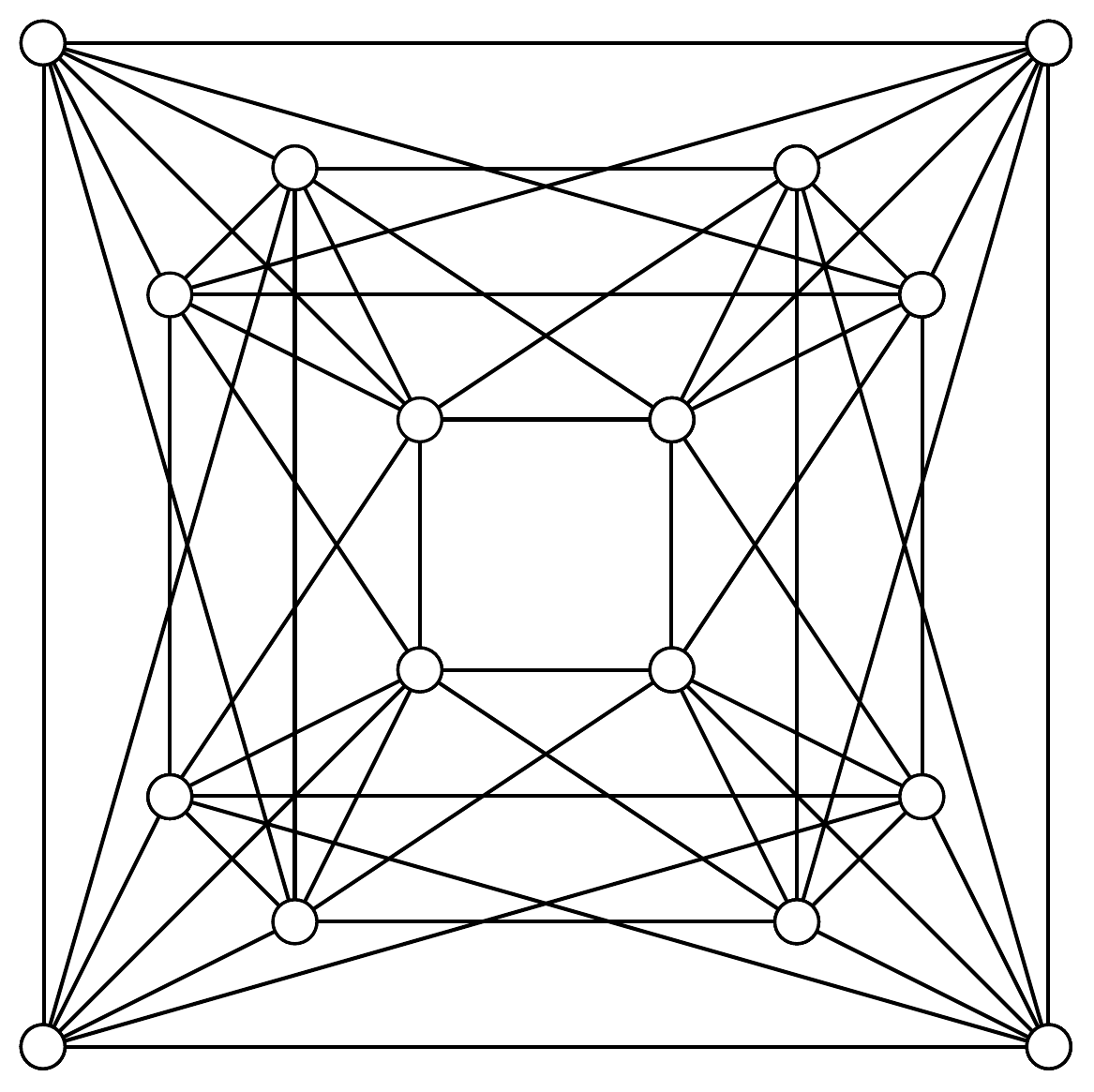}
	\caption{Graph of exclusivity $G_S$ associated to the CHSH Bell scenario. Intentionally, unlike the graph in Fig.~\ref{Fig1}, this graph has no colors or symbols in the nodes. Other sets of events produced in different KS scenarios also have $G_S$ as graph of exclusivity.}
	\label{Fig6}
\end{figure}

%%%%%%%%%%%%%%%%%%%%%%%%%%%%%%%%%%%%%%%%%%%%%%%%%%%%%%%%%%%%%%%%%%%

\subsection{Characterizing the quantum correlations for KS scenarios (and Bell scenarios with ideal measurements)}

%%%%%%%%%%%%%%%%%%%%%%%%%%%%%%%%%%%%%%%%%%%%%%%%%%%%%%%%%%%%%%%%%%%

In the previous subsections we have not made any reference to any specific scenario. We have simply shown that Assumptions~1 and~2 select the quantum set of assignments for any graph of exclusivity. In this subsection we will see which are the implications of this result for the sets of behaviors for specific scenarios.

Lemma~4 implies that the set of behaviors for any Bell or KS scenario~$S$ must be a {\em subset} of $\thb(G_S)$, where $G_S$ is the graph of exclusivity of the events of~$S$. For example, $G_S$ for the CHSH Bell scenario is shown in Fig.~\ref{Fig6}. 

In addition, the behaviors for~$S$ must satisfy constraints~(A), (B), and~(C) for~$S$ defined in Sec.~\ref{constraints}. It is easy to see how constraints~(A) and (B) exclude some elements of $\thb(G_S)$. However, the way constraint~(C) acts is more subtle.
Constraint~(C) demands that the mathematical representation of each event of~$S$ must be associated to the mathematical representation of the initial state, measurements, and outcomes that define the event. 
The following Lemma \cite{CSW14} helps us to understand how constraint (C) removes elements of $\thb(G_S)$ that satisfy constraints (A) and (B).

{\em Lemma~5.}
	Every element of $\thb(G_S)$ can be produced in a particular KS scenario $S_{\rm KS}$ defined by a set of two-outcome measurements whose graph of compatibility is isomorphic to $G_S$.

The proof follows from the definition of $\thb(G_S)$ in Eq.~(\ref{thb}). Lemma~5 tells us that every assignment in $\thb(G_S)$ can be produced in a suitable KS~scenario. 
However, in general, the KS (or Bell) scenario $S$ that we are interested in differs from $S_{\rm KS}$ and has a smaller number of measurements and (maximal) contexts. Then is when constraint~(C) becomes relevant. To illustrate how~(C) excludes elements of $\thb(G_S)$, we will focus on a particular type of scenario in which all the events are defined by the outcomes of two two-outcome measurements, i.e., all the events are of the type~$(x=a,y=b|\psi)$.

Any element of $\thb(G_S)$ must satisfy conditions (I'), (II'), and (III') in Sec.~\ref{cond}.
According to~(I'), the initial state of the system must be associated to a vector with unit norm $|\psi \rangle$. 
According to~(II'), the state of the system after two compatible measurements $x$ and $y$ with respective outcomes $a$ and $b$ on state $\psi$ must be associated to a vector with unit norm $|x=a,y=b|\psi \rangle$. Since $x$ and $y$ can be measured by measuring $x$, first, and measuring $y$ later on,
\begin{equation}
|x=a,y=b|\psi \rangle = |y=b|x=a, \psi \rangle,
\end{equation}
where $|x=a, \psi \rangle$ is the state after measuring $x$ and obtaining $a$ on state $\psi$. 

Also by~(II'), the state of the system after measuring $x$ and obtaining $a$ on state $\psi$ must be associated to another vector with unit norm, $|x=a|\psi \rangle$. To see how $|x=a,y=b|\psi \rangle$, $|x=a|\psi \rangle$, and $|\psi \rangle$ are related to each other we have to take into account the following.

Any additional measurement of $x$ after event $(x=a|\psi)$ does not disturb $x$. This is so because $x$ is an ideal measurement. Therefore, this additional measurement should not change event $(x=a|\psi)$. This implies that
	\begin{equation}
	|x=a, \psi \rangle = N^{(x)}_a E^{(x)}_a |\psi \rangle,
	\end{equation}
	where $N^{(x)}_a$ is a normalization constant and $E^{(x)}_a$ is a projector (i.e., a linear transformation from the vector space ${\cal V}$ to itself such that whenever it is applied twice gives the same result as if it were applied once) associated to the act of measuring $x$ and obtaining $a$.
	Similarly,
	\begin{equation}
	|y=b|x=a, \psi \rangle = N^{(y)}_b E^{(y)}_b |x=a, \psi \rangle,
	\end{equation}
	where $N^{(y)}_b$ is a normalization constant and $E^{(y)}_b$ is a projector associated to the act of measuring $y$ and obtaining $b$.

(II')~says that, for fixed $x$, $\{ |x=a | \psi \rangle \}$ must be a set of orthogonal unit vectors. Therefore,
	\begin{equation}
	E^{(x)}_{a} E^{(x)}_{a'} = \delta_{a,a'} E^{(x)}_{a}.
	\end{equation}
	Similarly, for fixed $x$ and $y$, $\{ |x=a, y=b | \psi \rangle \}$ must be a set of orthogonal unit vectors. Therefore,
	\begin{equation}
	E^{(y)}_{b} E^{(y)}_{b'} = \delta_{b,b'} E^{(y)}_{b}.
	\end{equation}

(III') and (A) imply that, for fixed $x$,
	\begin{equation}
	\sum_{a \in A} |\langle x=a | \psi | \psi \rangle|^2 =1.
	\end{equation}
	Therefore,
	\begin{equation}
	\sum_{a \in A} E^{(x)}_{a} =\id.
	\end{equation}
	Similarly, for fixed $x,y$,
	\begin{equation}
	\sum_{a \in A, b\in B} |\langle x=a, y=b | \psi | \psi \rangle|^2 =1.
	\end{equation}
	Therefore,
	\begin{equation}
	\sum_{b\in B} E^{(y)}_{b}=\id.
	\end{equation}

Finally, in Bell and KS scenarios, the order in which $x$ and $y$ are performed must be irrelevant. Therefore,
	\begin{equation}
	| x=a, y=b | \psi \rangle = |y=b, x=a | \psi \rangle, 
	\end{equation}
	which implies that
	\begin{equation}
	[E^{(x)}_{a}, E^{(y)}_{b}]=0\; \forall a,b,
	\end{equation}
	whenever $x$ and $y$ are compatible.
	
Therefore, event $(x=a|\psi)$ is associated to projector $E^{(x)}_{a}$ applied to vector $|\psi\rangle$, event $(y=b|\psi)$ to $E^{(y)}_{b}$ applied to $|\psi\rangle$, event $(x=a,y=b|\psi)$
to $E^{(x)}_{a}$ and $E^{(y)}_{b})$ (which commute) applied to $|\psi\rangle$, and $P(x=a,y=b|\psi)=|\langle \psi' | \psi \rangle |^2$, where $|\psi'\rangle = N_i E^{(x)}_{a} E^{(y)}_{b} |\psi\rangle$, where $N_i$ is a normalization constant. The fact that each scenario $S$ only has a limited number of measurements and that each event of $S$ can only be obtained using some specific projectors implies that not all the elements of $\thb(G_S)$ satisfying constraints (A) and (B) can be attained.

This example can be easily extended to any other type of scenario and then we can see that, for any Bell or KS scenario~$S$, conditions (I'), (II'), and (III'), which characterize the elements of $\thb(G_S)$, and constraints (A), (B), and (C) for $S$ imply conditions~(I), (II), and (III) for $S$ (as defined in Sec.~\ref{qc}). This finishes the proof of Theorem~1.

%%%%%%%%%%%%%%%%%%%%%%%%%%%%%%%%%%%%%%%%%%%%%%%%%%%%%%%%%%%%%%%%%%%

\section{Conclusions}
\label{Sec5}

%%%%%%%%%%%%%%%%%%%%%%%%%%%%%%%%%%%%%%%%%%%%%%%%%%%%%%%%%%%%%%%%%%%

\subsection{Characterizing quantum correlations from simple assumptions}

%%%%%%%%%%%%%%%%%%%%%%%%%%%%%%%%%%%%%%%%%%%%%%%%%%%%%%%%%%%%%%%%%%%

Here we have proven that, for Bell and for KS scenarios, no physical theory satisfying Assumptions~1 (there is a nonempty set of behaviors for any KS scenario) and~2 (there is a statistically independent joint realization of any two KS~experiments) can provide correlations different than those of~QT. QT produces {\em all} correlations satisfying these assumptions. 

In a nutshell, the proof runs as follows. Assumption~1 forces the theory to assign behaviors to any KS (and therefore Bell) scenario. Lemmas~1 and~2 state that the EP should hold for any KS and bipartite Bell scenario. Assumption~2 assures that there is a statistically independent joint realization of any two KS (including Bell) experiments. This joint realization can be seen as a single KS experiment or as a single bipartite Bell experiment and, therefore, their behaviors must satisfy the EP. Behaviors assign probabilities to the vertices of graphs of exclusivity. Therefore, assignments of probabilities to the graphs of exclusivity should satisfy the EP. Lemmas~4 and~5 show that $\thb(G)$ is, for every graph $G$, the largest set of assignments satisfying Assumptions~1 and~2. Therefore, $\thb(G)$ is the set of potential assignments of probability for any set of events that have $G$ as graph of exclusivity. The last part of the proof consists of showing that, for each KS (or Bell) scenario, constraints (A), (B), and (C) remove the non quantum behaviors from $\thb(G)$. 
 
Assumptions~1 and~2 are almost inevitable in any physical theory. Therefore, the fact that every correlation that is possible under Assumptions~1 and~2 is actually possible according to QT can be taken as an indication that our universe lacks laws restricting correlations for certain experiments. This argument is further developed in Ref.~\cite{Cabello19}.

Theorem~1 provides a solution to the problem of what selects the quantum correlations for Bell and KS scenarios and why higher-than-quantum correlations are impossible. For example, it provides an explanation of why the non-quantum behaviors in the set of almost quantum behaviors for Bell scenarios \cite{NGHA15} are impossible: each of them belongs to $\thb(G_S)$, where $G_S$ is the graph of exclusivity of the events of the corresponding Bell scenario $S$, and satisfy constraints (A) and (B). However, they fail to satisfy~(C) \cite{Cabello19}.

%%%%%%%%%%%%%%%%%%%%%%%%%%%%%%%%%%%%%%%%%%%%%%%%%%%%%%%%%%%%%%%%%%%

\subsection{Different perspective on mathematically unpleasant aspects of quantum correlations}

%%%%%%%%%%%%%%%%%%%%%%%%%%%%%%%%%%%%%%%%%%%%%%%%%%%%%%%%%%%%%%%%%%%

Given a Bell or KS scenario $S$, whether or not $\mathbf{p}(S)$ is allowed by QT is, in general, undecidable \cite{Slofstra17}. 
In addition, the set of quantum correlations for a given bipartite Bell scenario $S$ is, in general, not closed \cite{Slofstra17,DPP19}.
In Ref.~\cite{NGHA15}, the authors ask the following question: 
%[T]he problem of whether or not a given behavior is quantum is, in general, (\ldots) undecidable. 
``Do we really believe that correlations in nature have these properties?''

The approach followed in this article provides a different perspective. Quantum correlations can be separated into correlations for different scenarios, but also can be separated into correlations for different graphs of exclusivity. Given a graph of exclusivity $G$, deciding whether or not $\mathbf{p}(G)$ is allowed by QT is the solution of a single semi-definite program \cite{GLS86,GLS88}. Moreover, the set of quantum probabilities for any graph of exclusivity~$G$ is closed \cite{GLS86,GLS88,Knuth94}. 

Therefore, quantum correlations have beautiful mathematical properties when they are separated into sets for different graphs of exclusivity: decidability within each set is simple and each of the sets is closed. The lack of these properties when quantum correlations are separated into sets for different scenarios can then be attributed to the separation in scenarios rather than taken as an indication that something is wrong with the present understanding of correlations in nature.

Another interesting observation is that while, for some particular Bell (or KS) scenarios, ${\cal P}(S)$ is conjectured to require a Hilbert space that is the tensor product one two infinite dimensional Hilbert spaces \cite{PV10}, we can prove that, for any graph $G$, ${\cal P}(G)$ can be produced with a Hilbert space of dimension $|V(G)|+1$.

%%%%%%%%%%%%%%%%%%%%%%%%%%%%%%%%%%%%%%%%%%%%%%%%%%%%%%%%%%%%%%%%%%%

\subsection{Implications for quantum information and computation}

%%%%%%%%%%%%%%%%%%%%%%%%%%%%%%%%%%%%%%%%%%%%%%%%%%%%%%%%%%%%%%%%%%%

Quantum correlations produced in Bell and KS experiments are behind fundamental applications in quantum information and computation such as device-independent secure communication \cite{Ekert91,BHK05,ABGMPS07}, randomness amplification \cite{Colbeck06,PAMBMMOHLMM10}, distributed computation \cite{BZPZ04} (for Bell experiments), and some forms of quantum computation \cite{AB09,Raussendorf13,HWVE14,DGBR15,RBDOB17,BGK18} (for KS experiments). Any of these applications is thus affected by the limitations that, according to QT, these correlations have. Understanding where do these limitations come from is therefore both a fundamental and a practical problem. 
Specifically, it is important to know whether hypothetical physical theories beyond QT would allow for correlations different than the ones allowed by QT. 

The result presented in this article allows us to change the rating of some protocols based on correlations. A protocol that was declared secure against adversaries limited by QT can now be declared secure against adversaries limited by resources satisfying Assumptions~1 and~2. This will be further elaborated elsewhere.
 
%%%%%%%%%%%%%%%%%%%%%%%%%%%%%%%%%%%%%%%%%%%%%%%%%%%%%%%%%%%%%%%%%%%

\begin{acknowledgments}
The author thanks S.~Abramsky, B.~Amaral, M.~Ara\'ujo, K.~Bharti, G.~Chiribella, C.~A.~Fuchs, P.~Grangier, M.~J.~W.~Hall, G.~Jaegger, M.~Kleinmann, J.~Korbicz, M.~P.~M\"uller, M.~Navascu\'es, P.~Perinotti, J.~R.~Portillo, M.~F.~Pusey, A.~Ribeiro~de~Carvalho, R.~Shack, K.~Svozil, M.~Terra~Cunha, H.~M.~Wiseman, Z.-P.~Xu, M.~\.Zukowski, and W.~H.~Zurek for comments, and C.~Budroni, E.~Cavalcanti, S.~L\'opez-Rosa, and A.~J.~L\'opez-Tarrida for extensive comments on the successive versions of the manuscript. This work has been supported by the Foundational Questions Institute (FQXi) Grant~FQXi-RFP-1608 and has been partially financed by the Ministry of Science, Innovation and Universities (MICIU) Grant No.~FIS2017-89609-P with FEDER funds, the Conserjer\'{\i}a de Conocimiento, Investigaci\'on y Universidad, Junta de Andaluc\'{\i}a and European Regional Development Fund (ERDF) Grant No.~SOMM17/6105/UGR, and the Knut and Alice Wallenberg Foundation project ``Photonic Quantum Information.'' 
\end{acknowledgments}

%%%%%%%%%%%%%%%%%%%%%%%%%%%%%%%%%%%%%%%%%%%%%%%%%%%%%%%%%%%%%%%%%%%

\appendix

%%%%%%%%%%%%%%%%%%%%%%%%%%%%%%%%%%%%%%%%%%%%%%%%%%%%%%%%%%%%%%%%%%%

\section{Proof of Lemma 1}
\label{AppB}

%%%%%%%%%%%%%%%%%%%%%%%%%%%%%%%%%%%%%%%%%%%%%%%%%%%%%%%%%%%%%%%%%%%

In KS scenarios, by definition, measurements are ideal. Here we prove that events produced by ideal measurements satisfy the EP.
That is, for any set $T$ of events produced by ideal measurements and such that every two events of $T$ are mutually exclusive, the sum of the probabilities of all the events of $T$ is bounded by one.
%\end{lemma2}

%%%%%%%%%%%%%%%%%%%%%%%%%%%%%%%%%%%%%%%%%%%%%%%%%%%%%%%%%%%%%%%%%%%

The set of events can be written as $T=\{(x_i=a_i | \psi )\}^{n}_{i=1}$, where $\{ x_i\}^{n}_{i=1}$ is a set of ideal measurements and $\{ a_i\}^{n}_{i=1}$ their respective outcomes. By condition~(iii) in the definition of ideal measurement in Sec.~\ref{ide}, for each event $(x_i=a_i | \psi) \in T$, there is an ideal two-outcome coarse graining of $x_i$ in which all the outcomes different than $a_i$ are coarse grained into outcome $\overline{a_i}$. We can write this two-outcome coarse graining as
\begin{equation}
\label{D1}
\overline{x}_i = \{ (\overline{x}_i=a_i):=(x_i=a_i), (\overline{x}_i=\overline{a}_i) \}.
\end{equation}

Since every two events $(x_i=a_i | \psi)$ and $(x_j=a_j| \psi)$ in $T$ are mutually exclusive, there is a measurement $x_{ij}$ that produces both events so that each of the events is associated to a different outcome of $x_{ij}$. Consider the following three-outcome coarse graining of $x_{ij}$:
\begin{equation}
\label{cero}
\begin{split}
\overline{x}_{ij} =& \{ (\overline{x}_{ij}=a_i):=(x_i=a_i), (\overline{x}_{ij}=a_j):=(x_j=a_j), \\
& (\overline{x}_{ij}=\overline{a}_{ij}) \}.
\end{split}
\end{equation}

From the definition of $\overline{x}_{ij}$ in Eq.~(\ref{cero}), $\forall \psi$,
\begin{equation}
\label{tres}
P(\overline{x}_{ij}=a_j | \psi) = P(x_j=a_j | \psi).
\end{equation}
Therefore,
\begin{equation}
\label{dos}
P(\overline{x}_i=\overline{a}_i, \overline{x}_{ij}=a_j | \psi) = P(\overline{x}_i=\overline{a}_i, x_j=a_j | \psi),
\end{equation}
where $(\overline{x}_i=\overline{a}_i, \overline{x}_{ij}=a_j | \psi)$ is the event in which $\overline{x}_i$ and $\overline{x}_{ij}$ are measured on state $\psi$ and outcomes $\overline{a}_i$ and $a_j$ are obtained, respectively.
Since $\overline{x}_i$ is ideal and a coarse graining of $\overline{x}_{ij}$, then, $\forall \psi$,
\begin{equation}
\label{uno}
P(\overline{x}_{ij}=a_j | \psi) = P(\overline{x}_i=\overline{a}_i, \overline{x}_{ij}=a_j | \psi).
\end{equation}

Therefore, from Eqs.~(\ref{tres})--(\ref{uno}), 
\begin{equation}\label{eq:proofEprinciple}
P(\overline{x}_i=\overline{a}_i, x_j=a_j | \psi) = P(x_j=a_j | \psi).
\end{equation}

%%%%%%%%%%%%%%%%%%%%%%%%%%%%%%%%%%%%%%%%%%%%%%%%%%%%%%%%%%%%%%%%%%%
% Fig. 7
%%%%%%%%%%%%%%%%%%%%%%%%%%%%%%%%%%%%%%%%%%%%%%%%%%%%%%%%%%%%%%%%%%%

\begin{figure}[t!]
\vspace{4mm}
\includegraphics[width=8.4cm]{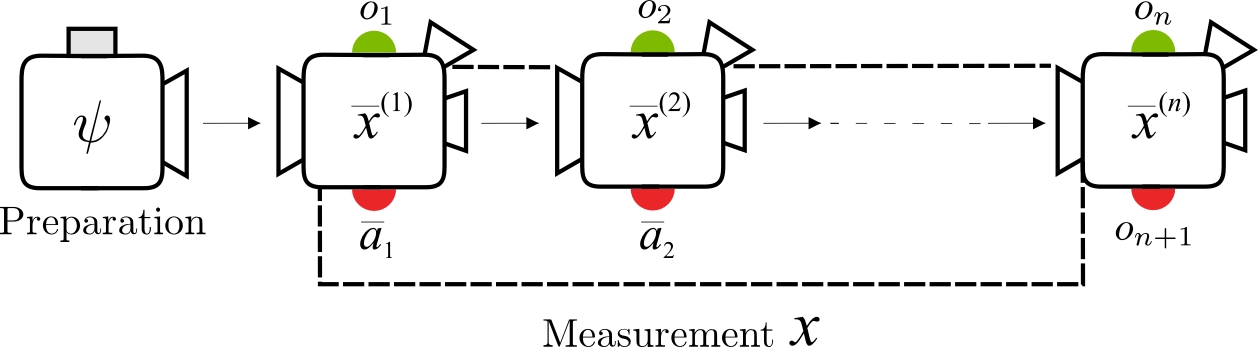}
\caption{Measurement $x$ used in the proof of Lemma~1.}
\label{Fig7}
\end{figure}

%%%%%%%%%%%%%%%%%%%%%%%%%%%%%%%%%%%%%%%%%%%%%%%%%%%%%%%%%%%%%%%%%%%

Now we can define the following $(n+1)$-outcome measurement $x$ with outcomes $ \{ o_i \}_{i=1}^{n+1}$: 
\begin{subequations}
\begin{align}
(x=o_1) &:= (\overline{x}^{(1)}=a_1), \\
(x=o_2) &:= (\overline{x}^{(1)}=\overline{a}_1,\overline{x}^{(2)}=a_2), \\ 
(x=o_i) &:= (\overline{x}^{(1)}=\overline{a}_1,\ldots,\overline{x}^{(i-1)}=\overline{a}_{i-1},\overline{x}^{(i)}=a_{i}),\\
(x=o_{n+1}) &:=(\overline{x}^{(1)}=\overline{a}_1,\ldots,\overline{x}^{(n-1)}=\overline{a}_{n-1},\overline{x}^{(n)}=\overline{a}_{n}),
\end{align}
\end{subequations}
where $i=3,\ldots ,n$.
Figure~\ref{Fig7} illustrates how $x$ is constructed. Then, $\forall \psi$,
%\begin{widetext}
\begin{eqnarray}
P(x=o_i|\psi) &=& P(\overline{x}^{(1)}=\overline{a}_1,\ldots, \overline{x}^{(i-1)}=\overline{a}_{i-1}, \overline{x}^{(i)}=a_i |\psi) \nonumber \\
&=& P(\overline{x}^{(i-1)}=\overline{a}_{i-1},\overline{x}^{(i)}=a_i|\psi_{i-2})\nonumber \\
&& \times P(\overline{x}^{(1)}=\overline{a}_1,\ldots,\overline{x}^{(i-2)}=\overline{a}_{i-2}|\psi),
\end{eqnarray}
where $\psi_j$ is the state after $(\overline{x}^{(1)}=\overline{a}_1,\ldots, \overline{x}^{(j)}=\overline{a}_j | \psi)$.

Taking into account that, by Eq.~(\ref{D1}), $(\overline{x}^{(i)}=a_i) = (x^{(i)}=a_i)$, then
\begin{eqnarray} 
P(x=o_i|\psi) &=& P(\overline{x}^{(i-1)}=\overline{a}_{i-1}, x^{(i)}=a_i|\psi_{i-2})\nonumber \\
&& \times P(\overline{x}^{(1)}=\overline{a}_1,\ldots, \overline{x}^{(i-2)}=\overline{a}_{i-2}|\psi).\hspace{20pt}
\end{eqnarray}
%\end{widetext}

Taking Eq.~(\ref{eq:proofEprinciple}) into account,
\begin{eqnarray}
P({\mathcal M}_{o_i}\psi|\psi)
&=& P(x^{(i)}=a_i|\psi_{i-2}) \nonumber \\
&& \times P(\overline{x}^{(1)}=\overline{a}_1,\ldots,\overline{x}^{(i-2)}=\overline{a}_{i-2}|\psi) \nonumber \\
&=& P(\overline{x}^{(1)}=\overline{a}_1,\ldots,\overline{x}^{(i-2)}=\overline{a}_{i-2},x^{(i)}=a_i |\psi).\hspace{24pt}
\label{nueve}
\end{eqnarray}
Applying a similar factorization, the right-hand side of Eq.~(\ref{nueve}) becomes
\begin{equation}
P(x^{(i)}=a_i|\psi_{i-3}) P(\overline{x}^{(1)}=\overline{a}_1, \ldots, \overline{x}^{(i-3)}=\overline{a}_{i-3} |\psi).
\end{equation}
Repeating this process, we obtain that, $\forall \psi$,
\begin{equation} 
\label{diez}
P(x=o_i|\psi)=P(x^{(i)}=a_i|\psi).
\end{equation}
By definition, $\{o_i\}^{n}_{i=1}$ are different outcomes of $x$. Therefore,
\begin{equation} 
\sum_{i=1}^{n} P(x=o_i |\psi) \leq 1.
\end{equation}
Hence, taking Eq.~(\ref{diez}) into account,
\begin{equation}
\sum_{i=1}^{n} P(x^{(i)}=a_i |\psi) \leq 1.
\end{equation}

%%%%%%%%%%%%%%%%%%%%%%%%%%%%%%%%%%%%%%%%%%%%%%%%%%%%%%%%%%%%%%%%%%%

\section{Proof of Lemma 2}
\label{AppA}

%%%%%%%%%%%%%%%%%%%%%%%%%%%%%%%%%%%%%%%%%%%%%%%%%%%%%%%%%%%%%%%%%%%

Behaviors in Bell scenarios must satisfy normalization and the nonsignaling principle (i.e., nondisturbance). In the following, we will prove that, for bipartite Bell scenarios, the set of behaviors satisfying the EP (applied to a single copy) is equal to the set of behaviors that satisfy normalization and the nonsignaling principle. 

Using the same notation used in Sec.~\ref{corre}, the events of any bipartite Bell scenario can be written as $(ab|xy)$, where $x \in X$ is Alice's measurement, $y \in Y$ is Bob's, $a \in A$ is the outcome of $x$, and $b \in B$ is the outcome of $y$. For fixed $x$, $y$, and $y'$, the following set of events:
\begin{equation}
\{(ab|xy) : b \in B\} \cup \{(a'b'|xy') : a' \in A \setminus a, b' \in B \}
\end{equation} 
is such that every pair contains mutually exclusive events. Therefore, the EP implies that
\begin{equation}
\sum_{b \in B} P(ab|xy) + \sum_{a' \in A \setminus a, b' \in B} P(a'b'|xy') \leq 1,
\end{equation} 
which, taking into account that, by normalization,
\begin{equation}
\sum_{a \in A, b' \in B} P(ab'|xy')=1,
\end{equation}
implies
\begin{equation}
\label{ine}
\sum_{b \in B} P(ab|xy) \leq \sum_{b' \in B} P(ab'|xy')
\end{equation}
for arbitrary $y$ and $y'$. By symmetry, the equality in (\ref{ine}) must hold. Therefore,
\begin{equation}
\label{ns1}
\forall a \in A, x \in X, \forall y, y' \in Y \; \sum_{b \in B} P(ab|xy) = \sum_{b' \in B} P(ab'|xy').
\end{equation}
Similarly, for fixed $x$, $x'$, and $y$, we obtain that
\begin{equation}
\label{ns2}
\forall b \in B, y \in Y, \forall x, x' \in X \; \sum_{a \in A} P(ab|xy) = \sum_{a' \in A} P(a'b|x'y).
\end{equation}
However, conditions~(\ref{ns1}) and (\ref{ns2}) characterize the set of behaviors that satisfy the nonsignaling principle.

%%%%%%%%%%%%%%%%%%%%%%%%%%%%%%%%%%%%%%%%%%%%%%%%%%%%%%%%%%%%%%%%%%%


\begin{thebibliography}{99}

\bibitem{Born26}
M. Born,
Zur Quantenmechanik der Sto{\ss}vorg\"ange,
\href{https://doi.org/10.1007/BF01397477}{Z. Physik \textbf{37}, 863 (1926)} 
[On the quantum mechanics of collisions,
in {\em Quantum Theory and Measurement}, 
edited by J. A. Wheeler and W. H. Zurek
(Princeton University Press, Princeton, NJ, 1983), p.~52.]

\bibitem{Bell64}
J. S. Bell,
On the Einstein Podolsky Rosen paradox,
\href{https://doi.org/10.1103/PhysicsPhysiqueFizika.1.195}{Physics \textbf{1}, 195 (1964).}

\bibitem{CHSH69}
J. F. Clauser, M. A. Horne, A. Shimony, and R. A. Holt,
Proposed Experiment to Test Local Hidden-Variable Theories,
\href{https://doi.org/10.1103/PhysRevLett.23.880}{Phys. Rev. Lett. \textbf{23}, 880 (1969).}

\bibitem{Specker60}
E. P. Specker,
Die Logik nicht gleichzeitig entscheidbarer Aussagen,
\href{https://doi.org/10.1111/j.1746-8361.1960.tb00422.x}{Dialectica \textbf{14}, 239 (1960)}
[English version: The logic of non-simultaneously decidable propositions, \href{http://arxiv.org/abs/1103.4537}{\eprint{arXiv:1103.4537}.}]

\bibitem{Bell66}
J. S. Bell,
On the problem of hidden Variables in quantum mechanics,
\href{https://doi.org/10.1103/RevModPhys.38.447}{Rev. Mod. Phys. \textbf{38}, 447 (1966).}

\bibitem{KS67}
S. Kochen and E. P. Specker,
The problem of hidden variables in quantum mechanics,
J. Math. Mech. \textbf{17}, 59 (1967).

\bibitem{BCPSW14}
N. Brunner, D. Cavalcanti, S. Pironio, V. Scarani, and S. Wehner,
Bell nonlocality,
\href{https://doi.org/10.1103/RevModPhys.86.419}{Rev. Mod. Phys. \textbf{86}, 419 (2014).}

\bibitem{Tsirelson80}
B. S. Cirel'son [Tsirelson],
Quantum generalizations of Bell's inequality,
\href{https://doi.org/10.1007/BF00417500}{Lett. Math. Phys. \textbf{4}, 93 (1980).}

\bibitem{PR94}
S. Popescu and D. Rohrlich,
Quantum nonlocality as an axiom,
\href{https://doi.org/10.1007/BF02058098}{Found. Phys. \textbf{24}, 379 (1994).}

\bibitem{vanDam99}
W. van Dam,
%{\em Nonlocality \& Communication Complexity},
Ph.D. thesis, University of Oxford, 1999.

\bibitem{LPSW07}
N. Linden, S. Popescu, A. J. Short, and A. Winter, 
Quantum Nonlocality and Beyond: Limits from Nonlocal Computation,
\href{https://doi.org/10.1103/PhysRevLett.99.180502}{Phys. Rev. Lett. \textbf{99}, 180502 (2007).}

\bibitem{PPKSWZ09}
M. Paw{\l}owski, T. Paterek, D. Kaszlikowski, V. Scarani, A. Winter, and M. \.{Z}ukowski,
Information causality as a physical principle,
\href{https://doi.org/10.1038/nature08400}{Nature (London) \textbf{461}, 1101 (2009).}

\bibitem{NW09}
M. Navascu\'es and H. Wunderlich,
A glance beyond the quantum model,
\href{https://doi.org/10.1098/rspa.2009.0453}{Proc. Royal Soc. A \textbf{466}, 881 (2009).}

\bibitem{NGHA15}
M. Navascu\'es, Y. Guryanova, M. J. Hoban, and A. Ac\'{\i}n,
Almost quantum correlations,
\href{https://doi.org/10.1038/ncomms7288}{Nat. Commun. \textbf{6}, 6288 (2015).}

\bibitem{KCBS08}
A. A. Klyachko, M. A. Can, S. Binicio\u{g}lu, and A. S. Shumovsky,
Simple Test for Hidden Variables in Spin-1 Systems,
\href{https://doi.org/10.1103/PhysRevLett.101.020403}{Phys. Rev. Lett. \textbf{101}, 020403 (2008).}

\bibitem{Cabello08}
A. Cabello,
Experimentally Testable State-Independent Quantum Contextuality,
\href{https://doi.org/10.1103/PhysRevLett.101.210401}{Phys. Rev. Lett. \textbf{101}, 210401 (2008).}

\bibitem{CSW10}
A. Cabello, S. Severini, and A. Winter,
(Non-)Contextuality of physical theories as an axiom,
\href{https://arxiv.org/abs/1010.2163}{\eprint{arXiv:1010.2163}.}

\bibitem{Cabello13}
A. Cabello,
Simple Explanation of the Quantum Violation of a Fundamental Inequality,
\href{https://doi.org/10.1103/PhysRevLett.110.060402}{Phys. Rev. Lett. \textbf{110}, 060402 (2013).}

\bibitem{CSW14}
A. Cabello, S. Severini, and A. Winter,
Graph-Theoretic Approach to Quantum Correlations,
\href{https://doi.org/10.1103/PhysRevLett.112.040401}{Phys. Rev. Lett. \textbf{112}, 040401 (2014).}

%%%%%%%%%%%%%%%%%%%%%%%%%%%%%%%%%%%%%%%%%%%%%%%%%%%%%%%%%%%%%%%%%%%
% Ideal measurements
%%%%%%%%%%%%%%%%%%%%%%%%%%%%%%%%%%%%%%%%%%%%%%%%%%%%%%%%%%%%%%%%%%%

\bibitem{CY14}
G. Chiribella and X. Yuan,
Measurement sharpness cuts nonlocality and contextuality in every physical theory,
\href{http://arxiv.org/abs/1404.3348}{\eprint{arXiv:1404.3348}.}

\bibitem{Kleinmann14}
M. Kleinmann,
Sequences of projective measurements in generalized probabilistic models,
\href{https://doi.org/10.1088/1751-8113/47/45/455304}{J. Phys. A: Math. Theor. \textbf{47}, 455304 (2014).}

\bibitem{CY16}
G. Chiribella and X. Yuan,
Bridging the gap between general probabilistic theories and the device-independent framework for nonlocality and contextuality,
\href{https://doi.org/10.1016/j.ic.2016.02.006}{Information and Computation
	\textbf{250}, 15 (2016).}

%%%%%%%%%%%%%%%%%%%%%%%%%%%%%%%%%%%%%%%%%%%%%%%%%%%%%%%%%%%%%%%%%%%
% For which measurements outcome noncontextuality is justified
%%%%%%%%%%%%%%%%%%%%%%%%%%%%%%%%%%%%%%%%%%%%%%%%%%%%%%%%%%%%%%%%%%%

\bibitem{Spekkens14}
R. W. Spekkens,
The status of determinism in proofs of the impossibility of a noncontextual model of quantum theory,
\href{https://doi.org/10.1007/s10701-014-9833-x}{Found. Phys. \textbf{44}, 1125 (2014).}

%%%%%%%%%%%%%%%%%%%%%%%%%%%%%%%%%%%%%%%%%%%%%%%%%%%%%%%%%%%%%%%%%%%
% Graph of compatibility
%%%%%%%%%%%%%%%%%%%%%%%%%%%%%%%%%%%%%%%%%%%%%%%%%%%%%%%%%%%%%%%%%%%

\bibitem{KRK12}
P. Kurzy\'{n}ski, R. Ramanathan, and D. Kaszlikowski,
Entropic Test of Quantum Contextuality,
\href{https://doi.org/10.1103/PhysRevLett.109.020404}{Phys. Rev. Lett. \textbf{109}, 020404 (2012).}

\bibitem{CDLP13}
A. Cabello, L. E. Danielsen, A. J. L\'opez-Tarrida, and J. R. Portillo,
Basic exclusivity graphs in quantum correlations,
\href{https://doi.org/10.1103/PhysRevA.88.032104}{Phys. Rev. A \textbf{88}, 032104 (2013).}

%%%%%%%%%%%%%%%%%%%%%%%%%%%%%%%%%%%%%%%%%%%%%%%%%%%%%%%%%%%%%%%%%%%
% Behaviors = empirical models = probability models
%%%%%%%%%%%%%%%%%%%%%%%%%%%%%%%%%%%%%%%%%%%%%%%%%%%%%%%%%%%%%%%%%%%

\bibitem{Tsirelson93}
B. S. Cirel'son [Tsirelson],
Some results and problems on quantum Bell-type inequalities,
Hadron. J. Suppl. \textbf{8}, 329 (1993).

\bibitem{AB11}
S. Abramsky and A. Brandenburger,
The sheaf-theoretic structure of non-locality and contextuality,
\href{https://doi.org/10.1088/1367-2630/13/11/113036}{New J. Phys. \textbf{13}, 113036 (2011).}

\bibitem{AH12}
S. Abramsky and L. Hardy,
Logical Bell inequalities,
\href{https://doi.org/10.1103/PhysRevA.85.062114}{Phys. Rev. A \textbf{85}, 062114 (2012).}

\bibitem{AFLS15}
A. Ac\'{\i}n, T. Fritz, A. Leverrier, and A. B. Sainz,
A combinatorial approach to nonlocality and contextuality,
\href{https://doi.org/10.1007/s00220-014-2260-1}{Comm. Math. Phys. \textbf{334}, 533 (2015).}

%%%%%%%%%%%%%%%%%%%%%%%%%%%%%%%%%%%%%%%%%%%%%%%%%%%%%%%%%%%%%%%%%%%
% Neumark's dilation theorem
%%%%%%%%%%%%%%%%%%%%%%%%%%%%%%%%%%%%%%%%%%%%%%%%%%%%%%%%%%%%%%%%%%%

\bibitem{Neumark40}
M. A. Neumark,
Self-adjoint extensions of the second kind of a symmetric operator,
Izv. Akad. Nauk S.S.S.R. S\'er. Mat. \textbf{4}, 53 (1940) (Russian with English summary);
%53--104
Spectral functions of a symmetric operator,
{\em ibid.} \textbf{4}, 277 (1940);
%277--318
On a representation of additive operator set functions,
C.R. (Dokl.) Acad. Sci. U.R.S.S. (N.S.) \textbf{41}, 359 (1943).
%359--361

\bibitem{Holevo80}
A. S. Holevo,
{\em Probabilistic and Statistical Aspects of Quantum Theory}
%(North-Holland, Amsterdam, 1982)
(Scuola Normale Superiore Pisa, Pisa, Italy, 2011), p.~55.
First published in Russian in 1980.

\bibitem{Peres95}
A. Peres,
{\em Quantum Theory: Concepts and Methods}
(Kluwer, New York, 1995), p.~285.

%%%%%%%%%%%%%%%%%%%%%%%%%%%%%%%%%%%%%%%%%%%%%%%%%%%%%%%%%%%%%%%%%%%
% The EP
%%%%%%%%%%%%%%%%%%%%%%%%%%%%%%%%%%%%%%%%%%%%%%%%%%%%%%%%%%%%%%%%%%%

\bibitem{Yan13}
B. Yan,
Quantum Correlations are Tightly Bound by the Exclusivity Principle,
\href{https://doi.org/10.1103/PhysRevLett.110.260406}{Phys. Rev. Lett. \textbf{110}, 260406 (2013).}

\bibitem{ATC14}
B. Amaral, M. Terra Cunha, and A. Cabello,
Exclusivity principle forbids sets of correlations larger than the quantum set,
\href{https://doi.org/10.1103/PhysRevA.89.030101}{Phys. Rev. A \textbf{89}, 030101(R) (2014).}

\bibitem{Cabello15}
A. Cabello,
Simple Explanation of the Quantum Limits of Genuine $n$-Body Nonlocality,
\href{https://doi.org/10.1103/PhysRevLett.114.220402}{Phys. Rev. Lett. \textbf{114}, 220402 (2015).}

\bibitem{Henson15}
J. Henson,
Bounding Quantum Contextuality with Lack of Third-Order Interference,
\href{https://doi.org/10.1103/PhysRevLett.114.220403}{Phys. Rev. Lett. \textbf{114}, 220403 (2015).}

\bibitem{CCKM19} 
G. Chiribella, A. Cabello, M. Kleinmann, and M. P. M\"uller,
General Bayesian theories and the emergence of the exclusivity principle,
\href{https://arxiv.org/abs/1901.11412}{\eprint{arXiv:1901.11412}.}

\bibitem{FSABCLA13}
T. Fritz, A. B. Sainz, R. Augusiak, J. Bohr Brask, R. Chaves, A. Leverrier, and A. Ac\'{\i}n,
Local orthogonality: A multipartite principle for correlations,
\href{https://doi.org/10.1038/ncomms3263}{Nat. Commun. \textbf{4}, 2263 (2013).}

%%%%%%%%%%%%%%%%%%%%%%%%%%%%%%%%%%%%%%%%%%%%%%%%%%%%%%%%%%%%%%%%%%%
% Graph theory
%%%%%%%%%%%%%%%%%%%%%%%%%%%%%%%%%%%%%%%%%%%%%%%%%%%%%%%%%%%%%%%%%%%

\bibitem{GLS86}
M. Gr\"otschel, L. Lov\'asz, and A. Schrijver,
Relaxations of vertex packing,
\href{https://doi.org/10.1016/0095-8956(86)90087-0}{J. Combin. Theory B \textbf{40}, 330 (1986).}

\bibitem{GLS88}
M. Gr\"otschel, L. Lov\'asz, and A. Schrijver,
{\em Geometric Algorithms and Combinatorial Optimization}
(Springer, Berlin, 1988).

\bibitem{Knuth94}
D. E. Knuth,
The sandwich theorem,
\href{http://www.combinatorics.org/ojs/index.php/eljc/article/view/v1i1a1}{The Electronic Journal of Combinatorics No.\ 1, A1, 1 (1994).}

\bibitem{BH03}
T. Bohman and R. Holzman, 
A nontrivial lower bound on the Shannon capacities of the complement of odd cycles,
\href{https://doi.org/10.1109/TIT.2002.808128}{IEEE~Trans. Inf. Theor. \textbf{49}, 721 (2003).}

\bibitem{Schwenk74}
A. J. Schwenk,
Computing the characteristic polynomial of a graph,
in
{\em Graphs Combinatorics}, 
Lecture Notes in Mathematics, vol.~406,
edited by R. Bary and F. Harary (Springer-Verlag, Berlin, 1974), p.~153.

%%%%%%%%%%%%%%%%%%%%%%%%%%%%%%%%%%%%%%%%%%%%%%%%%%%%%%%%%%%%%%%%%%%
% Conclusions
%%%%%%%%%%%%%%%%%%%%%%%%%%%%%%%%%%%%%%%%%%%%%%%%%%%%%%%%%%%%%%%%%%%

\bibitem{Cabello19}
A. Cabello,
The problem of quantum correlations and the totalitarian principle,
\href{https://doi.org/10.1098/rsta.2019.0136}{Phil. Trans. R. Soc. A \textbf{377}, 2019.0136 (2019).}

%%%%%%%%%%%%%%%%%%%%%%%%%%%%%%%%%%%%%%%%%%%%%%%%%%%%%%%%%%%%%%%%%%%
% Recent problems on quantum correlations for Bell scenarios
%%%%%%%%%%%%%%%%%%%%%%%%%%%%%%%%%%%%%%%%%%%%%%%%%%%%%%%%%%%%%%%%%%%

\bibitem{Slofstra17}
W. Slofstra,
The set of quantum correlations is not closed,
\href{https://doi.org/10.1017/fmp.2018.3}{Forum Math., Pi \textbf{7}, E1 (2019).}

\bibitem{DPP19}
K. Dykema, V. I. Paulsen, and J. Prakash,
Non-closure of the set of quantum correlations via graphs,
\href{https://doi.org/10.1007/s00220-019-03301-1}{Comm. Math. Phys. \textbf{365}, 1125 (2019).}

\bibitem{PV10}
K. F. P\'al and T. V\'ertesi,
Maximal violation of a bipartite three-setting, two-outcome Bell inequality using infinite-dimensional quantum systems,
\href{https://doi.org/10.1103/PhysRevA.82.022116}{Phys. Rev. A \textbf{82}, 022116 (2010).}

%%%%%%%%%%%%%%%%%%%%%%%%%%%%%%%%%%%%%%%%%%%%%%%%%%%%%%%%%%%%%%%%%%%
% Implications for quantum information
%%%%%%%%%%%%%%%%%%%%%%%%%%%%%%%%%%%%%%%%%%%%%%%%%%%%%%%%%%%%%%%%%%%

\bibitem{Ekert91}
A. K. Ekert,
Quantum Cryptography Based on Bell's Theorem,
\href{https://doi.org/10.1103/PhysRevLett.67.661}{Phys. Rev. Lett. \textbf{67}, 661 (1991).}

\bibitem{BHK05}
J. Barrett, L. Hardy, and A. Kent,
No Signaling and Quantum Key Distribution,
\href{https://doi.org/10.1103/PhysRevLett.95.010503}{Phys. Rev. Lett. \textbf{95}, 010503 (2005).}

\bibitem{ABGMPS07}
A. Ac\'{\i}n, N. Brunner, N. Gisin, S. Massar, S. Pironio, and V. Scarani,
Device-Independent Security of Quantum Cryptography against Collective Attacks,
\href{https://doi.org/10.1103/PhysRevLett.114.220403}{Phys. Rev. Lett. \textbf{98}, 230501 (2007).}

\bibitem{Colbeck06}
R. Colbeck,
Quantum and relativistic protocols for secure multi-party computation,
%Ph.D. thesis, University of Cambridge, 2006; 
\href{http://arxiv.org/abs/0911.3814 }{\eprint{arXiv:0911.3814}.}

\bibitem{PAMBMMOHLMM10}
S. Pironio, A. Ac\'{\i}n, S. Massar, A. Boyer de la Giroday,
D. N. Matsukevich, P. Maunz, S. Olmschenk, D. Hayes, L. Luo, T. A. Manning, and C. Monroe,
Random numbers certified by Bell's theorem,
\href{https://doi.org/10.1038/nature09008}{Nature (London) \textbf{464}, 1021 (2010).}

\bibitem{BZPZ04}
\v{C}. Brukner, M. \.{Z}ukowski, J.-W. Pan, and A. Zeilinger,
Bell's Inequalities and Quantum Communication Complexity,
\href{https://doi.org/10.1103/PhysRevLett.92.127901}{Phys. Rev. Lett. \textbf{92}, 127901 (2004).}

\bibitem{AB09}
J. Anders and D. E. Browne,
Computational Power of Correlations,
\href{https://doi.org/10.1103/PhysRevLett.102.050502}{Phys. Rev. Lett. \textbf{102}, 050502 (2009).}

\bibitem{Raussendorf13}
R. Raussendorf,
Contextuality in measurement-based quantum computation,
\href{https://doi.org/10.1103/PhysRevA.88.022322}{Phys. Rev. A \textbf{88}, 022322 (2013).}

\bibitem{HWVE14}
M. Howard, J. Wallman, V. Veitch, and J. Emerson,
Contextuality supplies the `magic' for quantum computation,
\href{https://doi.org/10.1038/nature13460}{Nature (London) \textbf{510}, 351 (2014).}

\bibitem{DGBR15}
N. Delfosse, P. A. Guerin, J. Bian, and R. Raussendorf,
Wigner Function Negativity and Contextuality in Quantum Computation on Rebits,
\href{https://doi.org/10.1103/PhysRevX.5.021003}{Phys. Rev. X \textbf{5}, 021003 (2015).}

\bibitem{RBDOB17}
R. Raussendorf, D. E. Browne, N. Delfosse, C. Okay, and J. Bermejo-Vega,
Contextuality as a Resource for Models of Quantum Computation with Qubits,
\href{https://doi.org/10.1103/PhysRevLett.119.120505}{Phys. Rev. Lett. \textbf{119}, 120505 (2017).}

\bibitem{BGK18}
S. Bravyi, D. Gosset, and R. K\"onig,
Quantum advantage with shallow circuits,
\href{https://10.1126/science.aar3106}{Science \textbf{362}, 308 (2018).}

%%%%%%%%%%%%%%%%%%%%%%%%%%%%%%%%%%%%%%%%%%%%%%%%%%%%%%%%%%%%%%%%%%%

\end{thebibliography}
\end{document}